\documentclass{article}
\usepackage[utf8]{inputenc}
\usepackage{lscape,epsfig,graphicx,subcaption}
\usepackage[section]{placeins}
\usepackage{makecell}
\usepackage{multirow}
\usepackage{siunitx,amssymb}
\usepackage{authblk}
\usepackage{color}
\usepackage{url}
\usepackage{overpic}
\usepackage {lineno}
\usepackage{textcomp}

\title{Design, waterproofing, and mass production of the 3-inch PMT frontend system of JUNO}

\author[8]{Jilei Xu\thanks{Corresponding authors. Email addresses:  xujl@ihep.ac.cn (Jilei Xu), hem@ihep.ac.cn (Miao He), cedric.cerna@in2p3.fr (C\'{e}dric Cerna), huangyb@gxu.edu.cn (Yongbo Huang).}}
\author[8]{Miao He$^*$}
\author[36]{C\'{e}dric Cerna$^*$}
\author[22]{Yongbo Huang$^*$}
\author[37]{Thomas Adam}
\author[56]{Shakeel Ahmad}
\author[56]{Rizwan Ahmed}
\author[16]{Fengpeng An}
\author[64]{Costas Andreopoulos}
\author[46]{Giuseppe Andronico}

% ...
\affil[8]{Institute of High Energy Physics, Beijing, China}
% ...
\affil[16]{Sun Yat-Sen University, Guangzhou, China}
% ...
\affil[22]{Guangxi University, Nanning, China}
% ...
\affil[36]{Univ. Bordeaux, CNRS, LP2I, UMR 5797, F-33170 Gradignan, F-33170 Gradignan, France}
\affil[37]{IPHC, Universit\'{e} de Strasbourg, CNRS/IN2P3, F-67037 Strasbourg, France}
% ...
\affil[46]{INFN Catania and Dipartimento di Fisica e Astronomia dell Universit\`{a} di Catania, Catania, Italy}
% ...
\affil[56]{Pakistan Institute of Nuclear Science and Technology, Islamabad, Pakistan}
% ...
\affil[64]{The University of Liverpool, Department of Physics, Oliver Lodge Laboratory, Oxford Str., Liverpool L69 7ZE, UK, United Kingdom}
% ...

\begin{document}

\maketitle 

\begin{abstract}
Over 25,600 3-inch photomultiplier tubes (PMTs) have been instrumented for the central detector of the Jiangmen Underground Neutrino Observatory. Each PMT is equipped with a high-voltage divider and a frontend cable with waterproof sealing. Groups of sixteen PMTs are connected to the underwater frontend readout electronics via specialized multi-channel waterproof connectors. This paper outlines the design and mass production processes for the high-voltage divider, the cable and connector, as well as the waterproof potting of the PMT bases. The results of the acceptance tests of all the integrated PMTs are also presented.

\end{abstract}


\section{Introduction}\label{sec.intro}

The Jiangmen Underground Neutrino Observatory (JUNO) is a particle physics experiment aiming for neutrino studies and exotic searches~\cite{An:2015jdp, Djurcic:2015vqa, JUNO:2022hxd}. The experiment started collecting physics data on August~26, 2025, and the initial detector performance results were documented in Ref.~\cite{JUNO:2025fpc}, with the first report on reactor neutrino oscillations measurement published in Ref.~\cite{JUNO:2025gmd}. The central detector~\cite{JUNO:2023ete} is made of a 20 kiloton liquid scintillator target surrounded by 17,612 20-inch~\cite{2022LPMTMass} and 25,600 3-inch photomultiplier tubes (PMTs)~\cite{Cao:2021wrq}, all of which are submerged in water at a depth of up to 43.5~m along with their frontend electronics.  

By virtue of their operation in single photon-counting mode at reactor antineutrino energies, the 3-inch PMT system provides a linear reference for calibrating any instrumental non-linearities in the charge response of the 20-inch PMTs~\cite{JUNO:2020xtj, Cabrera:2023dek}. The 3-inch PMTs also improve the energy resolution~\cite{JUNO:2024fdc}, enhance event reconstruction~\cite{Zhang:2024okq}, aid cosmic muon tracking~\cite{Yang:2022din}, and support several physics measurements, including proton decay searches~\cite{JUNO:2022qgr}, and the measurement of the solar oscillation parameters $\theta_{12}$ and $\Delta m^2_{21}$, for which they provide a semi-independent determination~\cite{JUNO:2022mxj}. 

\begin{figure}[!hbt]
  \centering
  \includegraphics[width=0.6\textwidth]{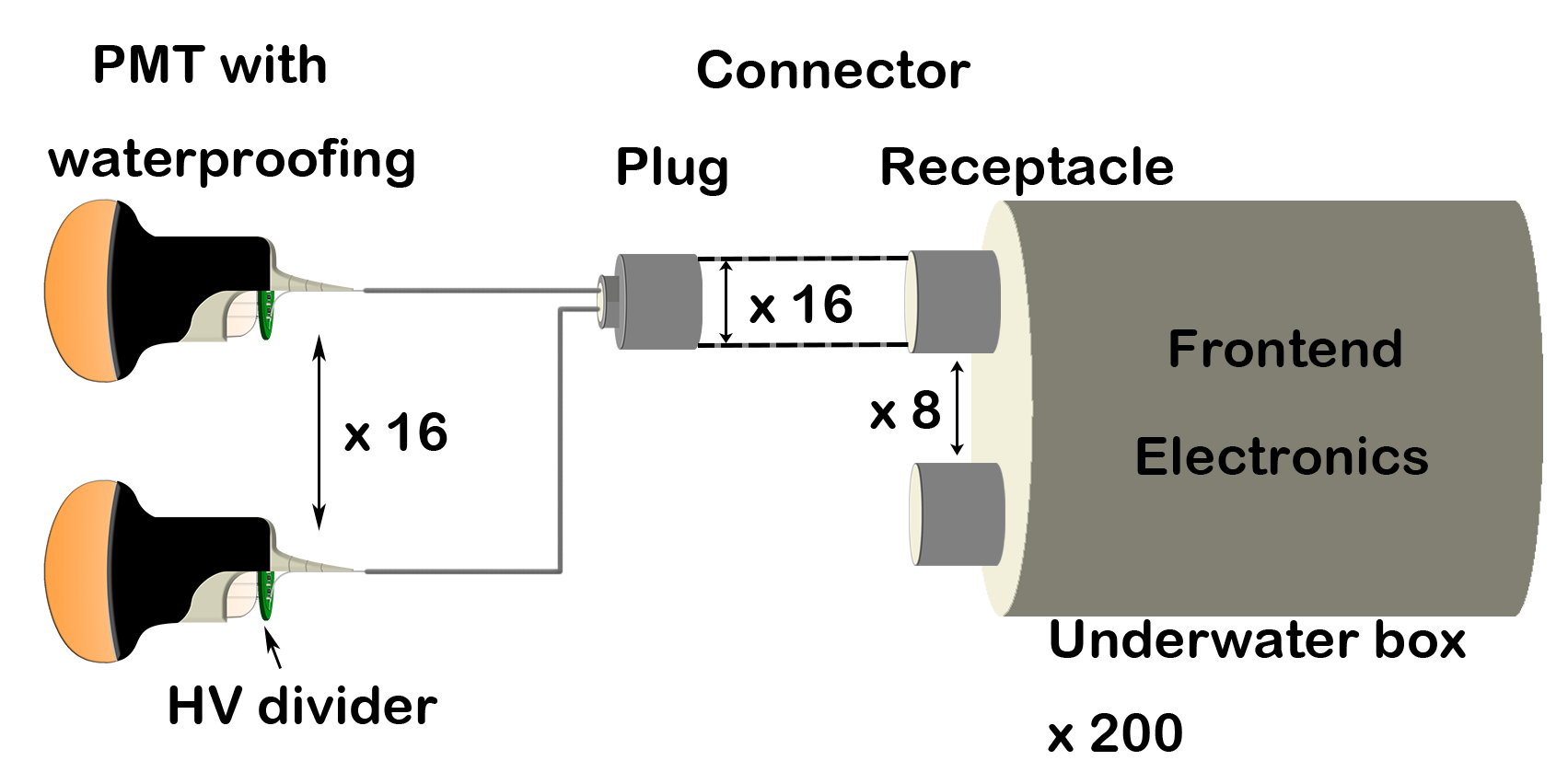}
  \caption{Diagram of a readout unit, consisting of 128 3-inch PMTs linked to underwater electronics through eight 16-channel connectors.}
  \label{fig:schematic}
\end{figure}

The system is divided into 200 readout units of 128 PMTs each. A diagram of one such unit is shown in Fig.~\ref{fig:schematic}. Each PMT is equipped with a high voltage (HV) divider and a coaxial cable that can be 5~m or 10~m in length and that transmits both the HV and the signal. The PMT, HV divider, and one end of the cable are soldered and waterproofed. The other ends of sixteen cables are connected to a multichannel connector, each consisting of a plug and receptacle pair. These connectors are linked to an underwater stainless steel box (UWB), housing the 128-channel frontend power supply and readout electronics. This paper provides details on the instrumentation of the 3-inch PMTs, encompassing the design and production of the HV divider, cable, and connector, as well as the waterproof potting and the acceptance tests. The production and characterization of 26,000 (including 400 spares) 3-inch PMTs were carried out at  Hainan Zhanchuang Photonics Technology Co., Ltd (HZC)~\cite{HZCnewfactory} and reported in Ref.~\cite{Cao:2021wrq}. The electronics design was outlined in Ref.~\cite{JUNO:2022hxd, JUNO:2020orn, Walker:2025xfa}.

The rest of this paper is structured as follows. In Sec.~\ref{sec.divider} and \ref{sec.connector}, the research and development, as well as the manufacturing process of the HV divider, cable, and connector, are presented. The waterproof potting is detailed in Sec.~\ref{sec.potting}. Section~\ref{sec.testing} covers the acceptance tests following potting. Section~\ref{sec.summary} provides a summary.
Unless otherwise specified, all references to PMTs in the subsequent sections pertain to the 3-inch PMTs.

\section{HV Divider}
\label{sec.divider}

\begin{figure}[ht]
\centering
\begin{subfigure}{.85\textwidth}
  \centering
  \includegraphics[width=.95\textwidth]{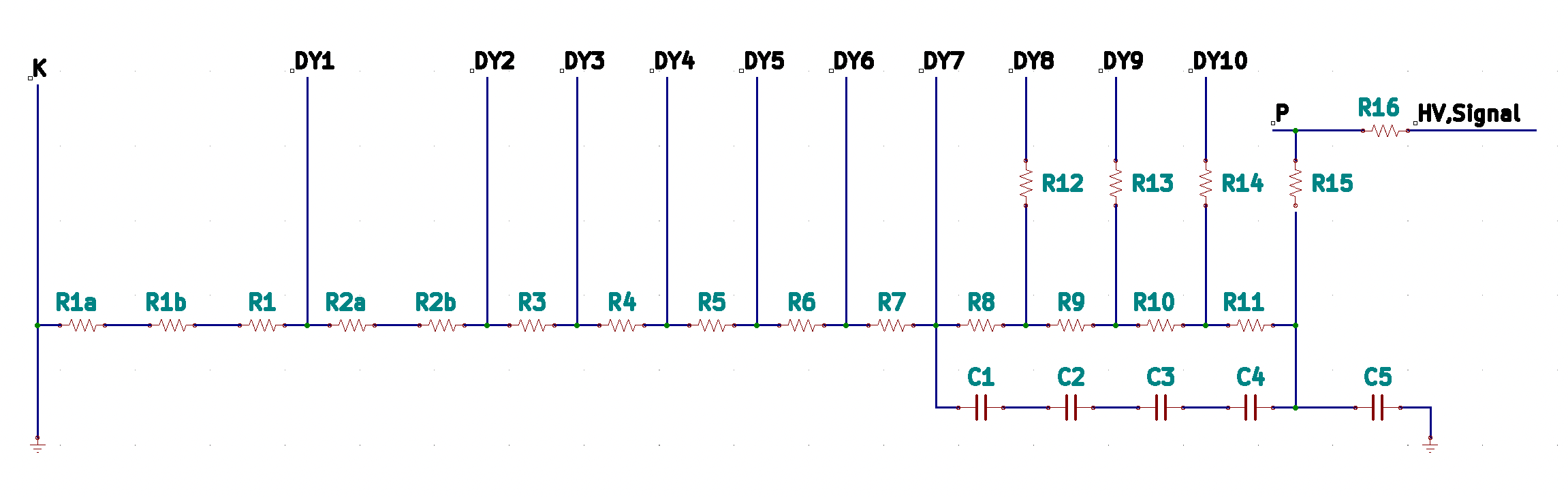}
  \caption{}
\end{subfigure}
\begin{subfigure}{.4\textwidth}
  \centering
  \includegraphics[width=.85\textwidth]{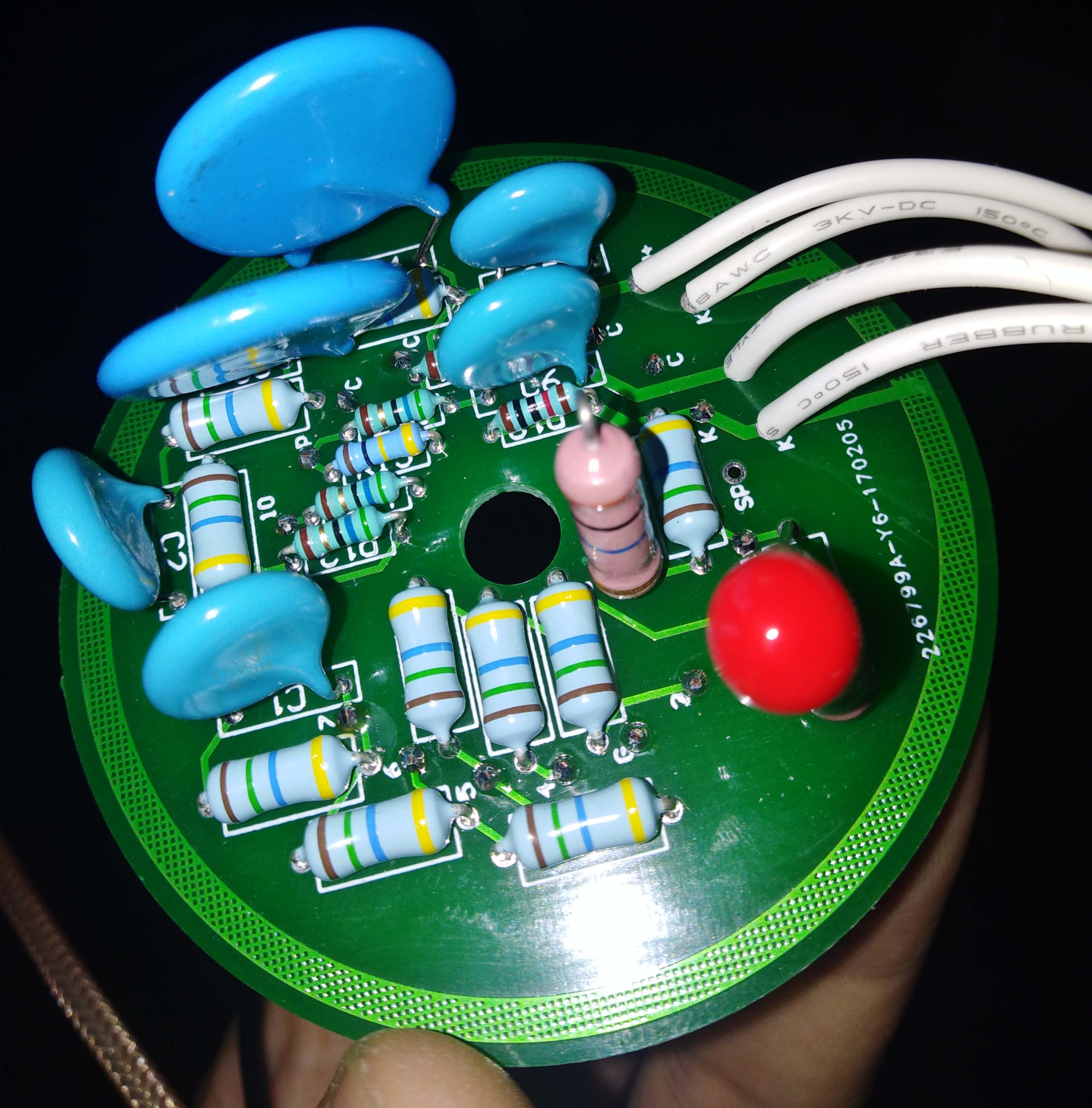}
  \caption{}
\end{subfigure}
\begin{subfigure}{.45\textwidth}
  \centering
  \includegraphics[width=.85\textwidth]{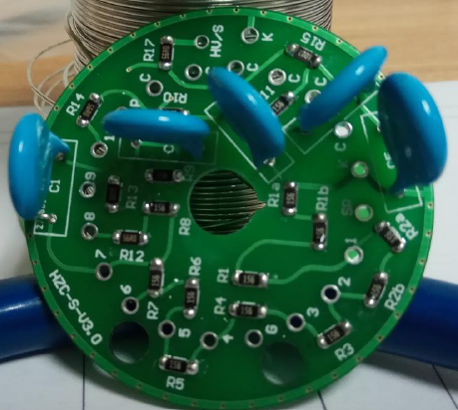}
  \caption{}
\end{subfigure}
\caption{(a) Schematic design of the 3-inch PMT HV divider. R1 to R11 are divider resistors, each with a resistance of 15~M$\Omega$. K denotes the cathode, DY1 to DY10 represent the ten dynode stages, and P indicates the anode. C1 to C4 are decoupling capacitors, each with a capacitance of 10~nF, used to maintain the signal linearity. R12 to R14 are damping resistors, each with a resistance of 56~$\Omega$, to prevent signal ringing. R15 (100~k$\Omega$) and C5 (4.7~nF) together form a low-pass filter. R16 is a matched resistor with a resistance of 56~$\Omega$. (b) Initial version of the divider with wire resistors. (c) Final version of the divider with surface-mounted resistors.}
\label{fig:divider}
\end{figure}

The HV divider follows a positive HV scheme. The HV ratio for the series of 11 dynodes is optimized for collection efficiency and timing measurements \cite{spmt-base, wanggang, caocy}. The resistor ratio for the first three dynodes is 3:2:1, and the other resistors are equal in value to the third one, 15~M$\Omega$ (Fig.~\ref{fig:divider}~(a)). The total resistance of 210~M$\Omega$ and a working HV generally less than 1,300\,V result in a gain of approximately $3\times10^6$ and a direct current (DC) less than 6.5~\textmu A. In consideration of safety and reliability requirements, the HV divider components include wire resistors, as depicted in Fig.~\ref{fig:divider}~(b), due to superior voltage and thermal tolerance. The main challenge lies in the dense packing of components on the small printed circuit board (PCB), which is designed to be soldered directly to the PMT pins and is constrained to have a similar size as the 48~mm diameter of the PMT glass bulb neck. Considering the re-evaluated requirements of the 3-inch PMT system for JUNO, the reliability requirement for the HV divider was slightly relaxed compared to the 20-inch PMT HV divider \cite{2022LPMTMass}, allowing for the use of surface-mounted components. The final design is illustrated in Fig.~\ref{fig:divider}~(c), where resistors of type 1206 with a maximum of 200~V and a 0.25~W power rating are selected, along with ceramic DC capacitors. The failure rate of the HV divider, as determined from reliability of individual components, is less than 0.05\% per year.

The HV divider is designed to achieve a maximum working HV of 2,000~V using a single layer of standard FR-4~\cite{webFR-4} PCB material with a thickness of 1.2~mm and a diameter of 47~mm, in compliance with the radioactivity requirements of JUNO~\cite{JUNO:2021kxb}. The PCB design guarantees a minimum clearance of 1.6~kV/mm between neighboring printed wires with varying voltages. Testing was performed on a few random samples of the final products, where HV dividers were subjected to several weeks of operation up to 2,400~V with satisfactory performance, and a maximum of 3,000~V was applied for a short period without failure. The flasher rate, which refers to the frequency of HV luminous microdischarges in the laboratory air of the HV divider, was verified to be less than 1~mHz, as detailed in Ref.~\cite{Yang:2020orr}. To mitigate potential risks of flasher occurrences during production and long-term operation, black polyurethane was used to cover the divider PCB after it has been soldered to the PMT pins, as will be discussed in Sec.~\ref{sec.potting}. A temperature cycling test was successfully conducted on 20 samples for 10 cycles without any failures, subjected to temperatures ranging from 5$^{\circ}$C to 55$^{\circ}$C for three hours per cycle. The signal quality coupled with the 3-inch PMT was further assessed in Ref.~\cite{3inchPMT-linan-2019,Wu:2022kke}, with additional results to be discussed in the following sections.

Viking~\cite{vikingcompany} surface mounting resistors of type 1206 and Murata~\cite{muratacompany} ceramic capacitors were chosen for the design.  In adherence to the JUNO divider PCB procedure~\cite{Yang:2020orr}, mass production was executed after a successful small batch production of 100 pieces, all of which passed testing. All PCBs were soldered at the factory following cleaning at the designated soldering points. Subsequently, a visual inspection of the soldering points was performed. After another round of cleaning, each PCB underwent a 30-second maximum HV test at 2,000V before being soldered to the PMT and the cable. In total, 26,500 divider boards were produced including 900 spares. Acceptance tests were conducted during integration with PMTs, as discussed in Sec.~\ref{subsec.potting}.

\section{Cable and Connector}
\label{sec.connector}

\subsection{Design} 

\begin{figure}[ht]
\centering
\begin{subfigure}{.55\textwidth}
  \centering
  \includegraphics[width=.95\textwidth]{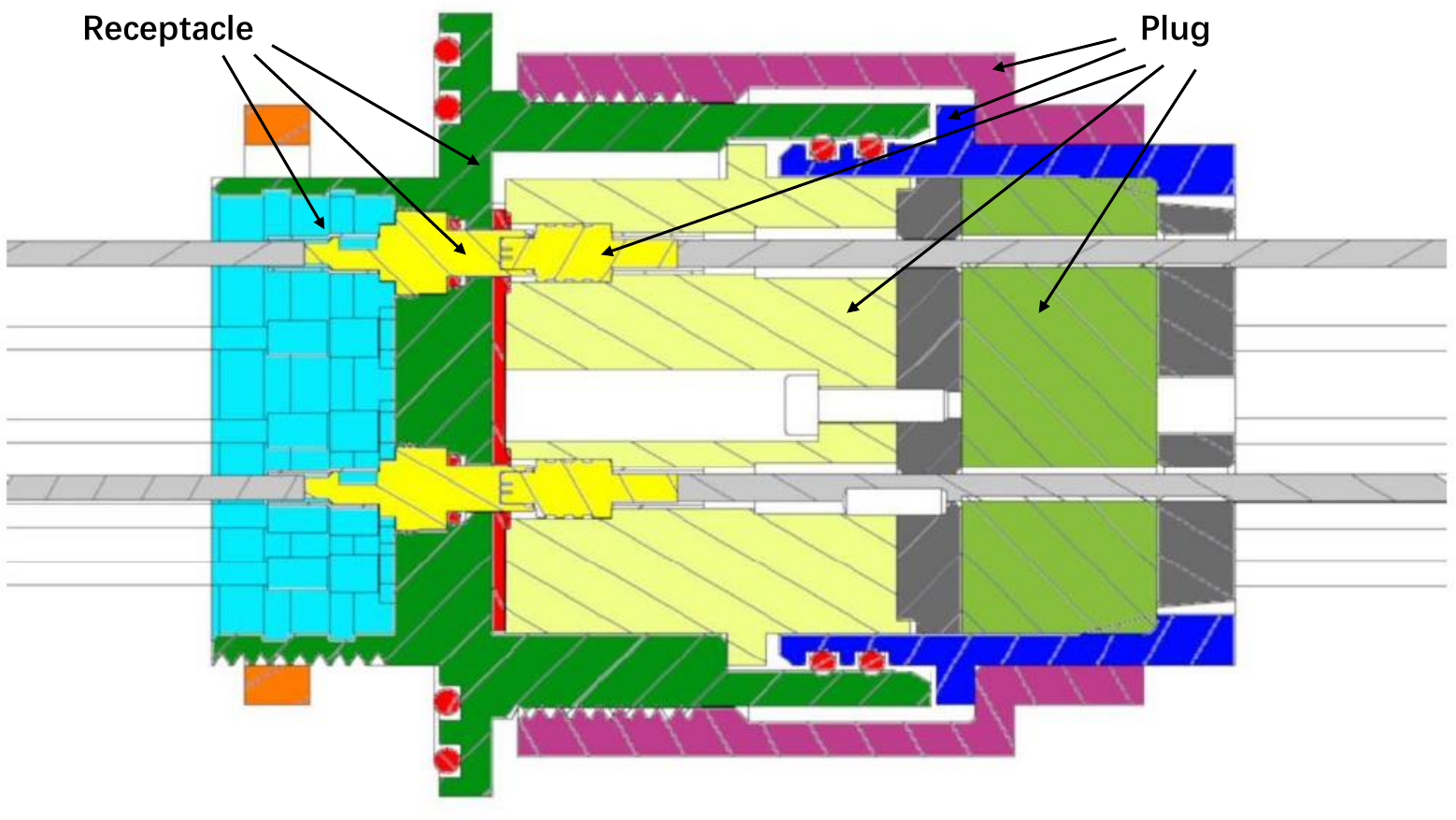}
  \caption{}
\end{subfigure}
\begin{subfigure}{.35\textwidth}
  \centering
  \includegraphics[width=.95\textwidth]{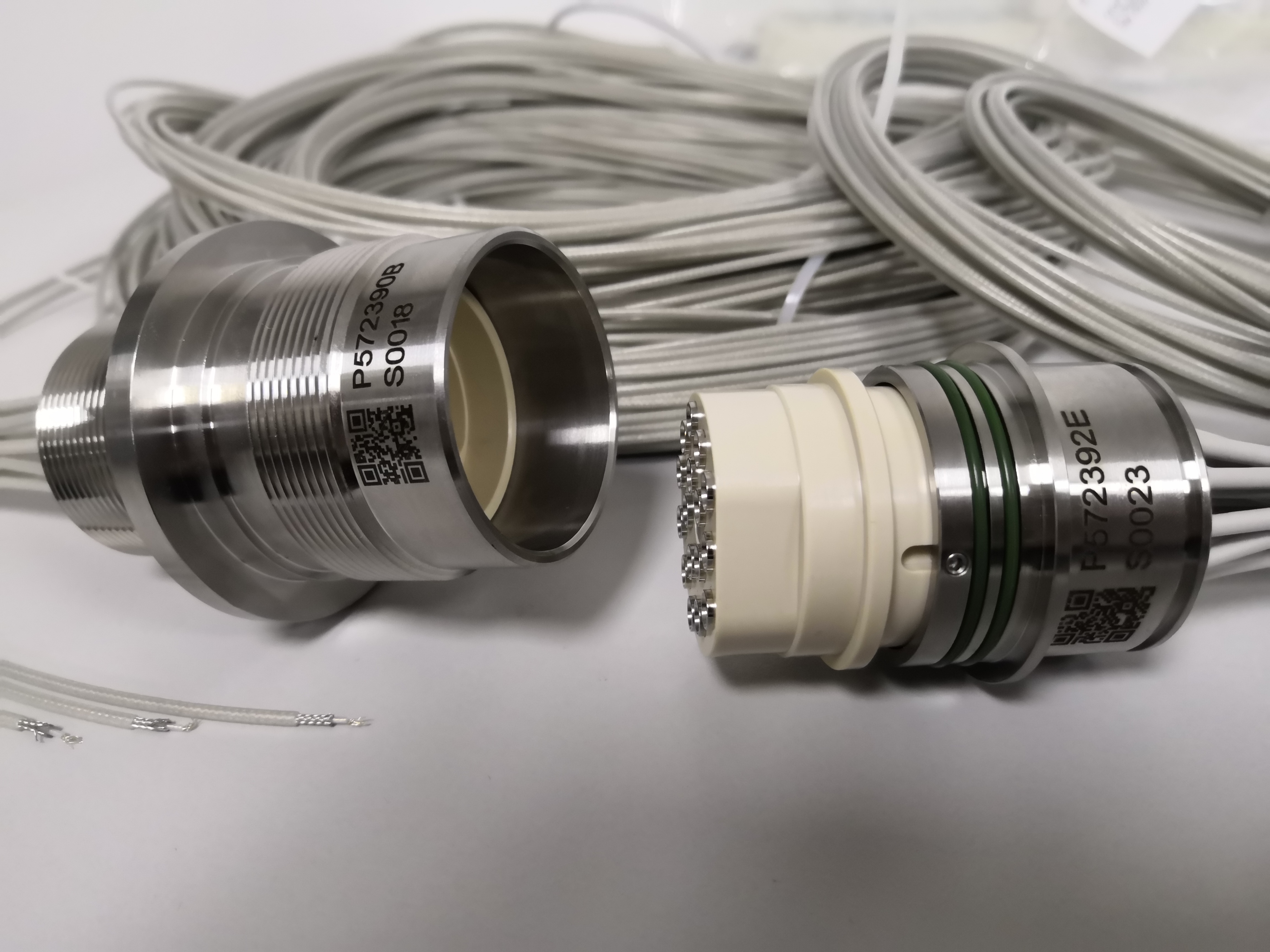}
  \caption{}
\end{subfigure}
\caption{(a) Schematic design of the 16-channel connector consisting of a plug (left part) and a receptacle (right part). The red dots and lines indicate water-resistant O-rings or gaskets. Further details are protected by intellectual property rights~\cite{patent-connector}. (b) A picture of a connector set with cables.}
\label{fig:connector}
\end{figure}

To reduce costs and facilitate the installation of the JUNO detector, a novel 16-channel waterproof connector and cable was designed in collaboration with AXON' company~\cite{AXON}. The connector's schematic design is shown in Fig.~\ref{fig:connector} and consists of two parts: a plug that is attached to 16 PMTs through frontend cables, and a receptacle that is mounted on the UWB. Waterproof sealing between the 16 cables is achieved through injection molding in the plug and the application of adhesive in the receptacle. A double-sealing mechanism relying on redundant axial and radial o-rings ensures tightness when the plug is connected to the receptacle.
Further design details and descriptions can be found in the Chinese patent~\cite{patent-connector}  held by AXON'.

The receptacle is approximately 25~cm in length and contains 16 specialized micro coaxial connector plug pins. These connector plugs are mounted on a special thermoplastic shell, which is subsequently inserted into the stainless steel shell of the receptacle. This design ensures that the cable grounding remains isolated from the receptacle shell, thus meeting JUNO's requirement of signal grounding to be separated from grounding of the JUNO mechanical structure. The end of the receptacle shell is sealed with epoxy, providing the receptacle with independent waterproof capability, regardless of whether the plug is connected. Eight receptacles are mounted on the lid of the UWB. Even if one plug fails, the receptacle is designed to prevent water from entering the UWB, thereby protecting the internal electronics.

A custom-made RG178 coaxial cable with a 50~$\Omega$ impedance was also developed. The outer layer of the cable is made of High-Density Polyethylene (HDPE) with a diameter of 2.1~mm, designed to prevent fluorescence and ensure long-term reliability in the ultra-pure water environment. To provide an extra layer of protection, the grounding layer was enhanced with a water-blocking powder based on sodium polyacrylate, a superabsorbent polymer developed by AXON'.
If the cable jacket is compromised, water would penetrate this layer, activating this barrier and preventing further water ingress. A water pressure test demonstrated that after one hour under 5~bar pressure, water infiltration was restricted to only 5~cm.

During the commissioning phase of the JUNO detector, a few weeks after the start of water filling, a significant number of channels experienced HV breakdown. The similar phenomenon was also observed in the SNO detector two decades ago \cite{SNO-det}. This issue was traced to air diffusing from the connectors toward the ultra-pure degassed water. This occurred because the air concentration within the connector exceeded that of the degassed water, leading to a drop in pressure within the connector assembly.  Since the insulation between the core and the ground of coaxial pins is provided by the residual air within the connector at atmospheric pressure, the breakdown voltage decreases as the pressure decreases according to Paschen's Law~\cite{Paschen}.  
To address this problem, a continuous flow of nitrogen gas was introduced into the water circulation system, which increase the overall gas concentration to levels higher than inside the connector. As a result, nitrogen gas diffuses from the water into the connector. This elevated the overall gas concentration to levels surpassing those inside the connector. Consequently, nitrogen gas diffused from the water into the connector, raising the pressure inside. 
This increase in pressure consequently raises the breakdown voltage, ensuring it exceeds the operational high voltage requirement. The 3-inch PMT system was integrated into the detector operation following the completion of liquid scintillator filling. The operating power of each PMT group is being continuously monitored and has remained stable. Further details and comprehensive analysis of this issue and its solution will be provided in an upcoming publication.

\subsection{Signal Performance}
\label{sec.signal}

\begin{figure}[ht]
\centering
\begin{subfigure}{.45\textwidth}
  \centering
  \includegraphics[width=.95\textwidth]{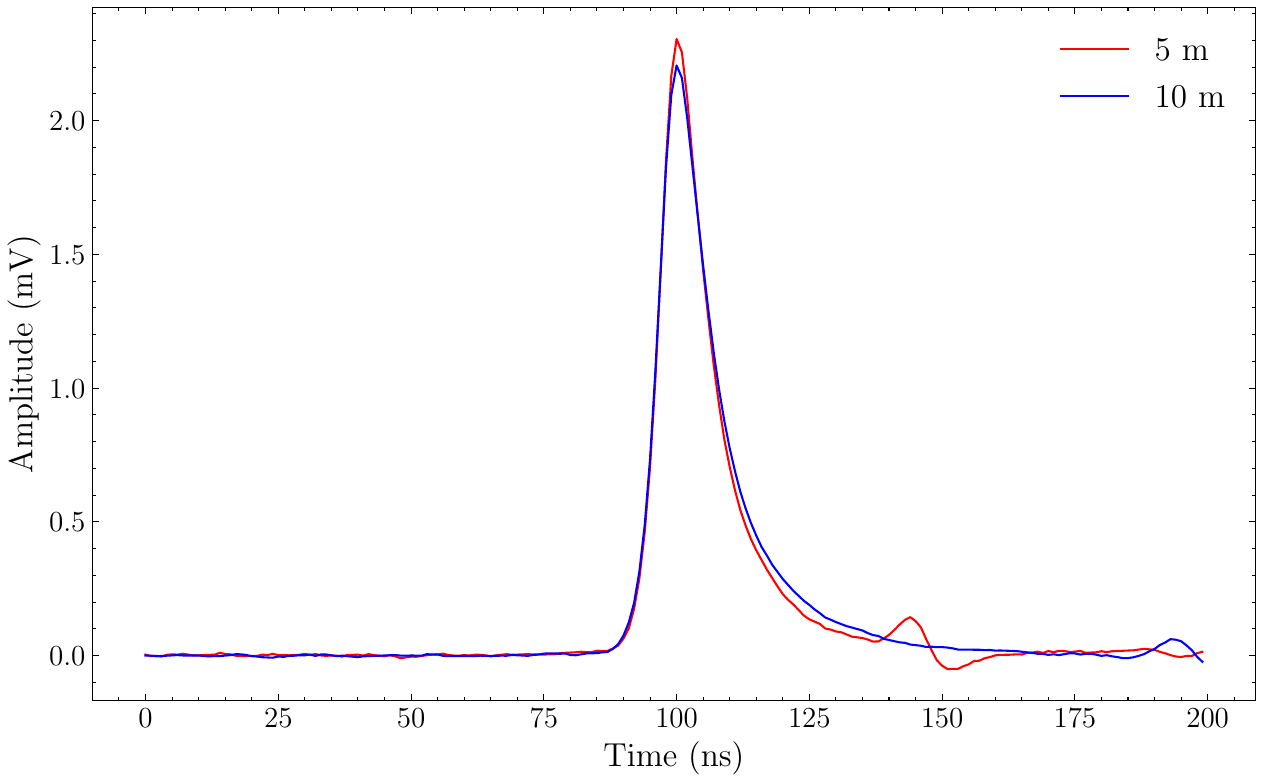}
  \caption{}
\end{subfigure}
\begin{subfigure}{.45\textwidth}
  \centering
  \includegraphics[width=.95\textwidth]{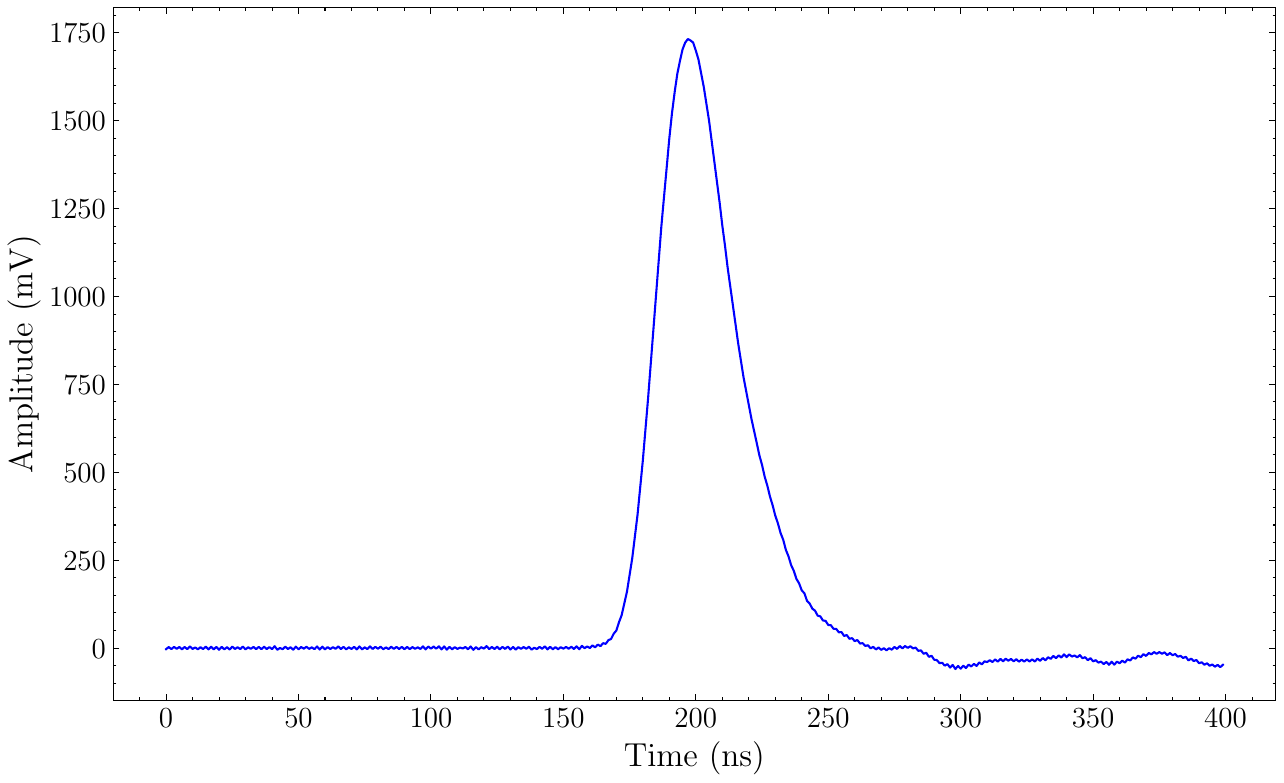}
  \caption{}
\end{subfigure}
\caption{(a) An average PMT signal for the single photoelectron using a 5~m (red) or 10~m (blue) cable. The reflection, resulting from a slight impedance mismatch in the prototype HV splitter, was addressed and resolved in the final version of the splitter~\cite{Walker:2025xfa}. (b) The average PMT signal with a large light intensity of approximately 4,000 photoelectrons. }
\label{fig:aveWaveforms}
\end{figure}

The quality of the PMT signal was examined using a prototype of the connector and cables, as detailed in Ref.~\cite{Wu:2022kke}. Typical signals for the single photoelectron (SPE) with a gain of 3$\times$10$^{6}$ are displayed in Fig.~\ref{fig:aveWaveforms}~(a). Small reflections following the main signal, attributed to a slight impedance mismatch not to the connector but in the hand-made HV-signal splitter, were observed. This issue was resolved in the final version of the splitter as reported in Ref.~\cite{Walker:2025xfa}. To compensate for the loss of PMT gain due to cable attenuation, an additional 23~V or 45~V voltage was applied for 5~m or 10~m cables, respectively. The observed amplitude, rise time, fall time, and Full Width at Half Maximum (FWHM) are listed in Tab.~\ref{Tab:SameGain}.

\begin{table}[htb]
\caption{Measured features of PMT SPE signals with cable length 5~m and 10~m at a gain of $3\times10^6$. Errors are statistical only.}
\label{Tab:SameGain}
\begin{center}
\begin{tabular}{ccc}
\hline\hline
Gain $3\times10^6$   & 5 m               & 10 m       \\ \hline
Amplitude (mV)    & $2.25\pm0.08$   & $2.16\pm0.08$   \\ \hline
Rise\ time (ns)  & $4.29\pm0.12$   & $4.57\pm0.12$   \\ \hline
Fall\ time (ns)   & $15.65\pm0.16$  & $18.68\pm0.14$  \\ \hline
FWHM (ns)        & $11.76\pm0.12$  & $12.98\pm0.12$  \\ 
\hline\hline
\end{tabular}
\end{center}
\end{table}

The measured PMT charge is not proportional to the number of incident photoelectrons due to both internal factors, such as the space charge effect, and external factors, such as the divider current~\cite{PMTbook-photonis}. Fractional loss of charge, known as nonlinearity, was observed in the JUNO 3-inch PMTs and reported in Ref.~\cite{Wu:2022gdk}. The nonlinearity remains under 5\% below 750~PEs but increases significantly above that due to the low divider current (less than 6.5~$\mu$A). An average signal of approximately 4,000 incident PEs obtained with high-intensity LED lights is shown in Fig.~\ref{fig:aveWaveforms}~(b). The integrated charge corresponds to 2,342 PEs when comparing with the single PE signal in Fig.~\ref{fig:aveWaveforms}~(a), which represents about 40\% charge nonlinearity. However, this nonlinearity is not an issue for JUNO, because the 3-inch PMTs operate in SPE regime for most physics cases.

The crosstalk for a complete assembly of cables, plugs and receptacles was assessed using an adapted network analyzer (Agilent E5062A)~\cite{agilentE5062A}. Considering two adjacent channels as the extreme case, in the frequency range of interest from 100~MHz to 350~MHz, 
an S-parameter analysis verified that the analog charge crosstalk between different channels is less than 0.3\%.

\subsection{Long Term Validation}

The JUNO 3-inch PMT system was designed to achieve a channel failure rate below 10\% over a 6-year period, equally allocated between the PMTs and their readout electronics. Under the assumption of a constant failure rate described by an exponential lifetime distribution, a 5\% failure probability over six years for the electronics corresponds to a Mean Time Between Failures (MTBF) of 117 years. In practice, this requirement is implemented by assuming an MTBF exceeding 120 years, which is equivalent to a Failure In Time (FIT) value below 951. This requirement applies directly to the cables and connectors that interface with the photomultipliers. Estimating such a low failure rate is challenging when dealing with a limited number of components over a short testing period. For instance, evaluating the reliability of two underwater connector plug and receptacle systems would necessitate a 40-year assessment.
In ultra-pure water, the primary cause of long-term failure is chemical degradation, which compromises the protective barriers against water ingress such as cable jackets and connector gaskets. The test duration can be accelerated by subjecting the system to increased temperatures in a controlled setting.  Compared to JUNO's operational water temperature of  $21^\circ$C, raising the temperature by an additional $10^\circ$C can expedite the testing process by a factor of approximately 2.5~\cite{IEC61709:2017}. Thus, a test conducted at $60^\circ$C for 12 months simulates the effect of 40 years of ageing at the normal operational temperature of $21^\circ$C.

To conduct the test, two pressurized water tanks were developed and equipped - one located in France and the other in Chile. These tanks allowed for the adjustment of water temperature and pressure from 1 to 10 bars to simulate various depths in JUNO. In the test conducted in France, four underwater connectors with 10-meter long cables were mounted on their receptacles. An electric potential ranging from 0 to 2~kV was applied to each cable, with the leakage current monitored at the micro-ampere level, allowing for the detection of any minor water leaks on the cable jacket or at the connector level. In Chile's test, five receptacles and three mechanical models were mounted on the UWB to evaluate long-term water leakage aging under conditions involving heat and pressure.

After the facilities were qualified, the following test campaigns were carried out:
\begin{itemize}
    \item A first period of 12 months between December 2020 and November 2021, with overpressure cycling between 5 to 10 bars and temperature varying between $55^\circ$C and $62^\circ$C in France;
    \item A second period of 4 months between December 2021 and March 2022, with overpressure between 1 to 2 bars, mimicking shallow water, and a temperature varying between $15^\circ$C and $20^\circ$C in France.
    \item Five receptacles with water temperature 88.7$^\circ$C and pressure from 4 to 9 bars, average 7.75 bars were tested in 2021 in Chile.  
\end{itemize}

During the two testing phases conducted in France, spanning over 11,000 hours, no leaking current was observed, and both the cable and connector functioned perfectly, achieving the required reliability with a FIT value smaller than 500. During the destructive test conducted in Chile, five receptacles were aged to the equivalent of over 120 years and exhibited no leaks, yielding a FIT value of 300.

\subsection{Production and Quality Inspection in Axon'}

Approximately 200~km of cables were produced and thoroughly tested, with 100\% of the insulation integrity verified through a 3,000~V spark test during the cable production process. Subsequently, the cable was cut to specific lengths, 10~m and 20~m for the plugs, and 25~cm for the receptacles.

\begin{figure}[!hbt]
  \centering
  \includegraphics[width=0.4\textwidth]{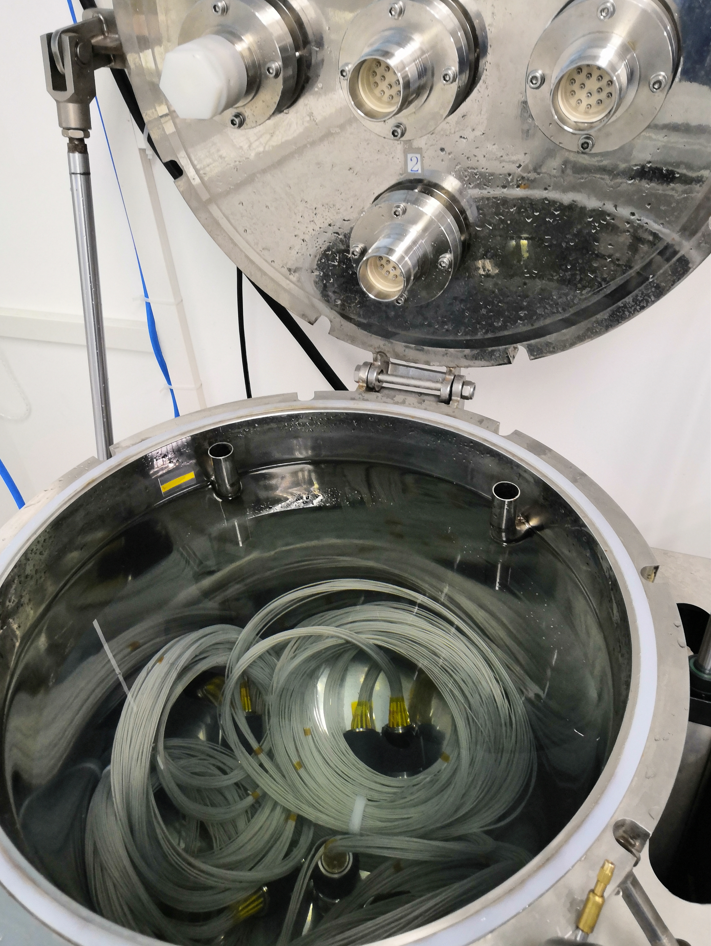}
  \includegraphics[width=0.4\textwidth]{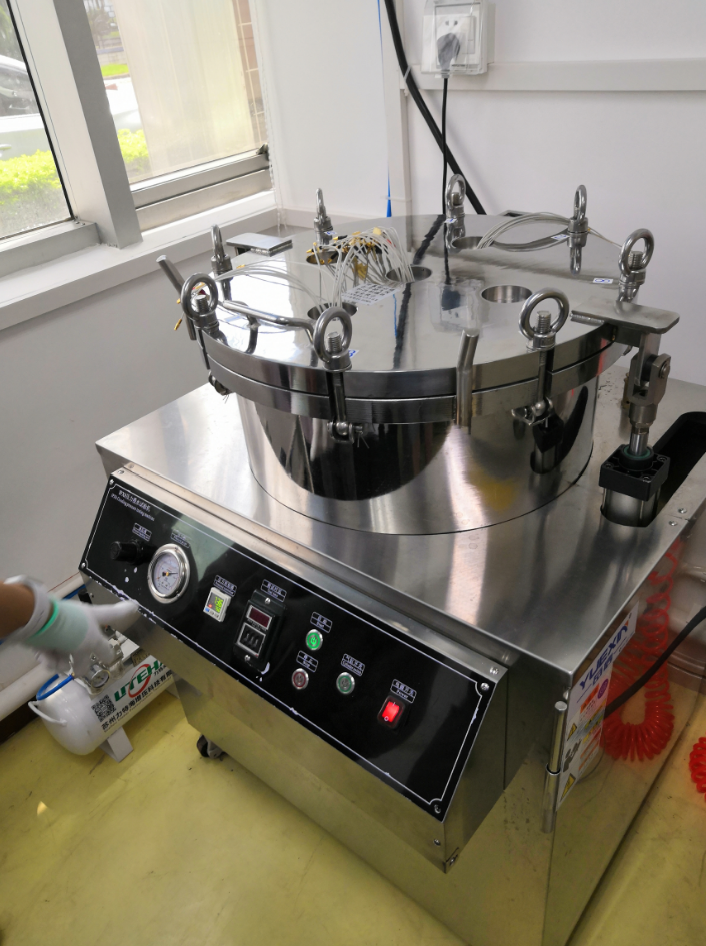}
  \caption{Leak test in a high-pressure water tank at AXON'. The dual-end plugs were sealed with fake receptacles and directly put into water. The receptacles were mounted on the lid of the tank. }
  \label{fig:waterTank}
\end{figure}

A bundle of 16 cables, either 10~m or 20~m long, were equipped with plugs at both ends and exposed to 1.8~kV voltage for 70 seconds to validate their operation. Cables that successfully passed this evaluation were enclosed by two imitation receptacles and immersed in an ultrapure water-filled stainless steel tank, as shown in Fig.~\ref{fig:waterTank}. Moisture test paper was included within the connector. The tank was closed, and the water pressure was raised to 0.75~MPa, sustained for two hours. Following removal from the water tank and drying, the effectiveness of water sealing was confirmed via a visual examination of the moisture test paper and a subsequent HV test. All cables and connectors underwent rigorous testing procedures. Those that met the specified criteria were then bisected at the midpoint to create the required 10~m and 5~m segments.

For each batch, the first and last receptacles were mounted on the lid of the water tank with the 16 pins to be submerged in water at a pressure of 0.75~MPa and sustained for two hours (refer to Fig.~\ref{fig:waterTank}).  
This testing procedure was classified as destructive due to the exposure of the pins to water. If the two receptacles mentioned above successfully passed this test, the remaining receptacles within the same batch proceeded to the subsequent stage. This stage involved connecting a micro coaxial connector to the opposite end of the cable, which links to the electronics in the UWB. In the case of a failure, all other receptacles in the batch underwent a similar underwater test, but this time sealed with a plug. Subsequently, all receptacles were subjected to the standard HV test at 1.8 kV for a duration of 70 seconds.

In total, 825 plugs of 10 meters in length and 825 plugs of 5 meters in length, along with 1,650 receptacles (including 50 spares), were produced and delivered to JUNO.

\subsection{Acceptance Tests by JUNO}

A warehouse designated for the JUNO 20-inch PMT mass testing~\cite{2022LPMTMass} was also used for the storage and acceptance tests of the 3-inch PMT cables and connectors. The initial step involved visual inspections to detect any visible damage or defects, with particular attention paid to scratches on the sealing surfaces of the connectors. Electrical tests followed, assessing the resistance and electrical conduction of the core wire and the braid grounding wire, as well as verifying the channel sequence from 1 to 16. Additionally, HV testing (1.6~kV, $>$ 5~minutes) was carried out to determine the cables' ability to withstand voltage stress and confirm their safe operation under normal conditions. Since only two receptacles from each batch underwent underwater tests at Axon', a supplementary leak test with SF$_{6}$ gas was conducted for 10\% of the receptacles. Each selected sample was sealed with a fake plug using an axial O-ring, evacuated to below 50~Pa, maintained for one minute, and then filled with SF$_{6}$ gas at a pressure about 25\% higher than standard atmospheric pressure. An SF$_{6}$ detector was used outside the receptacle to detect any leaked gas with a sensitivity of 10$^{-8}~$Pa$\cdot$m$^3$/s~\cite{Chu:2025rge}. No leakage was detected in any of the tested receptacles.

\begin{figure}[!hbt]
  \centering
  \includegraphics[width=0.4\textwidth]{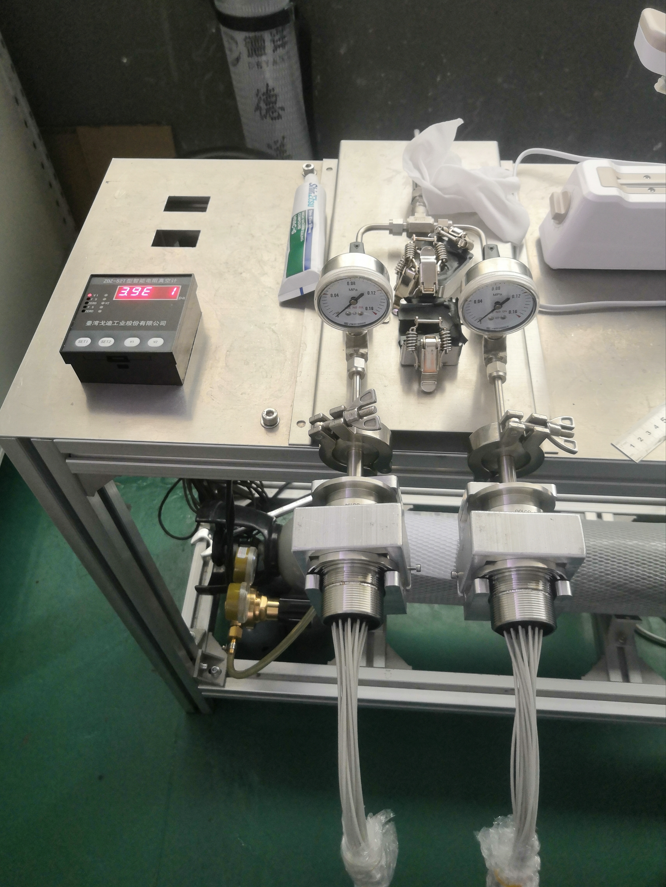}
  \caption{An SF$_{6}$-based leakage test facility during inspection. Two  receptacles undergo evacuation and are subsequently filled with SF$_{6}$ gas to perform the leakage test.}
  \label{fig:receptacleSF6test}
\end{figure}

\section{Integration and Waterproof Potting}
\label{sec.potting}

\subsection{PMT Grouping}

The 26,000 qualified and delivered bare PMTs~\cite{Cao:2021wrq} were assembled into groups of 16 according to their working HV at a gain of $3\times10^6$. Additionally, to minimize the risk of PMT implosion underwater, these PMTs were grouped according to their weights and installed at similar depths in the JUNO detector. 

\textbf{Grouping by HV}

\begin{figure}[!hbt]
  \centering
  \includegraphics[width=0.6\textwidth]{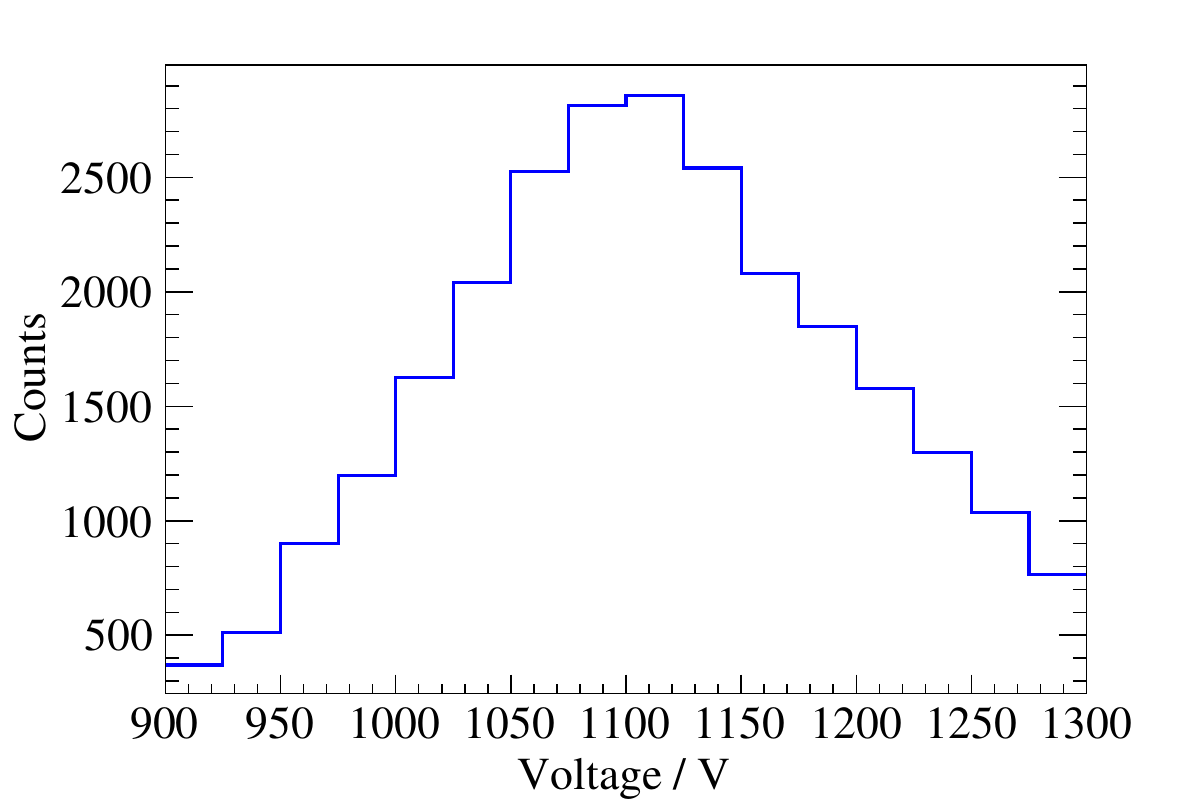}
  \caption{Distribution of working voltage at $3\times10^6$ gain of PMTs.}
  \label{fig:HV}
\end{figure}

The operating HV required to achieve a gain of $3\times10^6$ for each PMT was determined by the manufacturer, with the distribution shown in Fig.~\ref{fig:HV}. The voltage ranges from 900~V to 1,300~V. PMTs with working voltage outside of this range were excluded from mass production~\cite{Cao:2021wrq}. The voltages were categorized into 16 groups, each spanning 25~V, with a maximum gain difference of 15\% to 20\% within each group.

\textbf{Grouping by weight}

PMTs are made of quartz glass with an average thickness between 1~mm and 2~mm. During long-term operation in water at a 42~m depth, buckling failure may happen due to large external pressure. For an ideal spherical shell, the classical critical buckling pressure $p_b$ in MPa can be expressed as~\cite{buckling}

\begin{equation}\label{eq:bkl}
p_b = \frac{2E}{\sqrt{3(1-\nu^2)}}\frac{t^2}{r^2},
\end{equation}
where $\nu$ is the Poisson's ratio, a dimensionless quantity, $E$ is the tensile modulus expressed in MPa, and $r$
and $t$ are the radius and thickness of the glass shell in centimeter, respectively. Since $\nu$ and $E$ are constants for the PMT glass shell, and the shape (ratius $r$) is essentially fixed, the critical pressure is sensitive to the thickness of the glass and thus to the weight.

A dedicated study was made to find the dependence of the critical pressure on the weight of the PMT. Ten PMTs were randomly selected from an early production batch and their weights were measured. All pins on the bottom of the PMT were sealed with polyurethane and each PMT was individually put into a water tank. The water pressure was increased with a water pump at a speed of about 0.5~MPa/min. When the PMT glass imploded, the water pressure was recorded.

\begin{figure}[!hbt]
  \centering
  \includegraphics[width=0.6\textwidth]{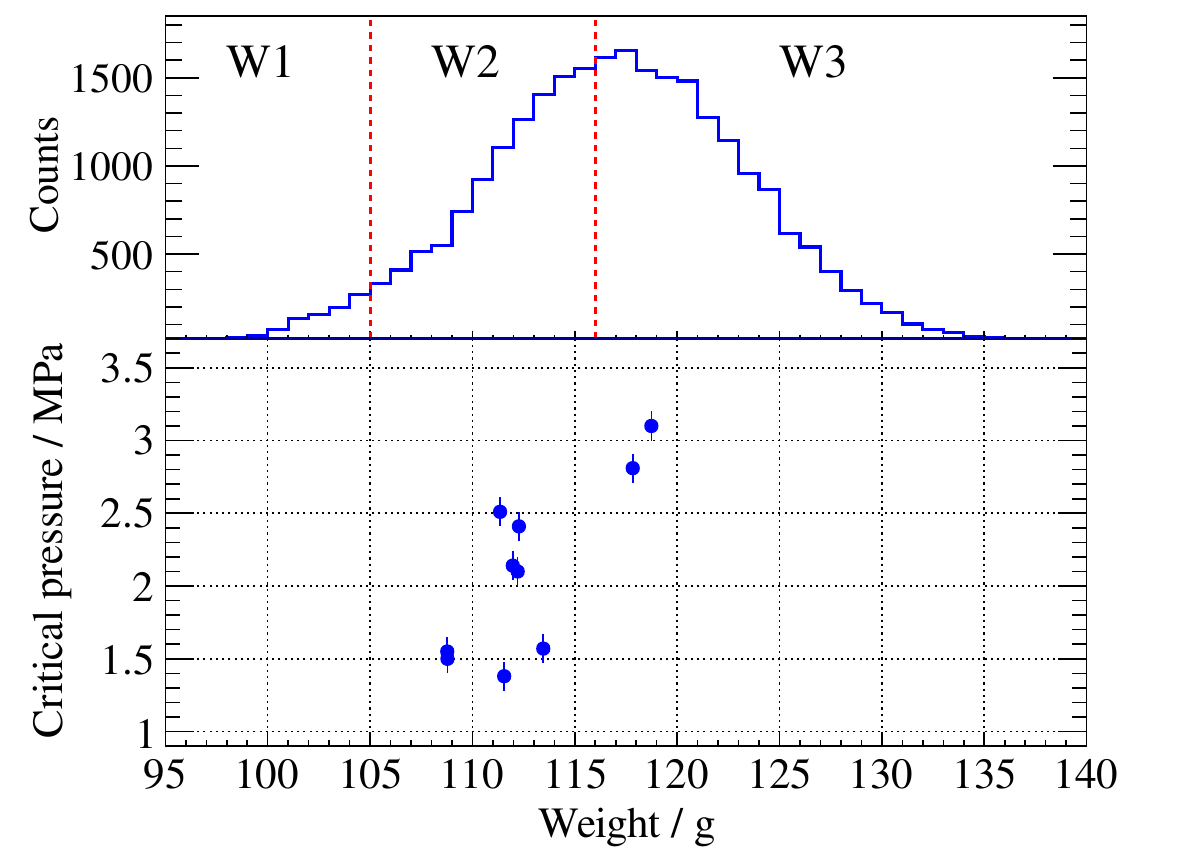}
  \caption{The top panel shows the measured weight distribution for all 26,000 PMTs. W1, W2, and W3 represent three groups of PMTs according to their weights. The blue points in the bottom panel are the measured critical pressures underwater as a function of the PMT weight.}
  \label{fig:pressure}
\end{figure}

Critical pressures in comparison to atmospheric pressure at varying weights of PMTs are presented in the lower panel of Fig.~\ref{fig:pressure}. A positive relationship is evident. However, there is a large spread from 1.5~MPa to 2.5~MPa in the 110~g to 115~g range, which can be attributed to variations in the glass shell thickness, as implosions typically initiate at points with the smallest thickness or highest stress. 
The obtained data was fitted with either a parabolic function as denoted in Eq.~\ref{eq:bkl} or a linear function, giving comparable fitting qualities. A critical pressure exceeding 1~MPa was observed for glass bulbs weighing over 105~g when extrapolating using either of the two fitting functions.

The weights of all 26,000 PMTs were measured and the resulting distribution is shown in the top panel of Fig.~\ref{fig:pressure}.
They were categorized into three classes based on their weights and installed at maximum water depths of 10~m, 25~m, and 42~m, respectively, as detailed in Tab.~\ref{tab:wgroup}. Only 3.8\% of PMTs fall into class W1 and were installed near the top of the JUNO detector at less than 10~m depth. About 40.5\% PMTs, belonging to class W2, were installed on the upper hemisphere of the JUNO detector at depths smaller than 25~m. The remaining PMTs, comprising more than half of the total, belong to class W3 and were installed mostly on the lower hemisphere. The critical pressure for W2 and W3 is three times greater than the anticipated water pressure. Although the safety margin is lower for W1, the proportion of PMTs in this category is minimal.

	\begin{table}
	\centering
	\caption{Classification of PMTs based on weight. The water pressure indicated in the final column is presented relative to atmospheric pressure.}
	\label{tab:wgroup}
	\begin{tabular}{ccccc}
	\hline
	\hline
		Class & Fraction & Weight & Installation depth & Water pressure in the leakage test \\
		\hline
        W1 & 3.8\% & [90, 105]~g & $<$10~m & 0.20~MPa \\
        W2 & 40.5\% & [105, 116]~g & $<$25~m & 0.40~MPa \\
        W3 & 55.7\% & [116, 150]~g & $<$42~m & 0.65~MPa \\
        \hline
        \hline
	\end{tabular}

	\end{table}

In total, 48 groups were formed, comprising three weight-based groups and 16 HV-based groups. During the potting process, PMTs were selected from groups that were as homogeneous as possible. The installation depth was determined based on the lightest PMT in each assembly.

\subsection{Waterproof PMT Potting}
\label{subsec.potting}

\begin{figure}[!hbt]
  \centering
  \includegraphics[width=0.7\textwidth]{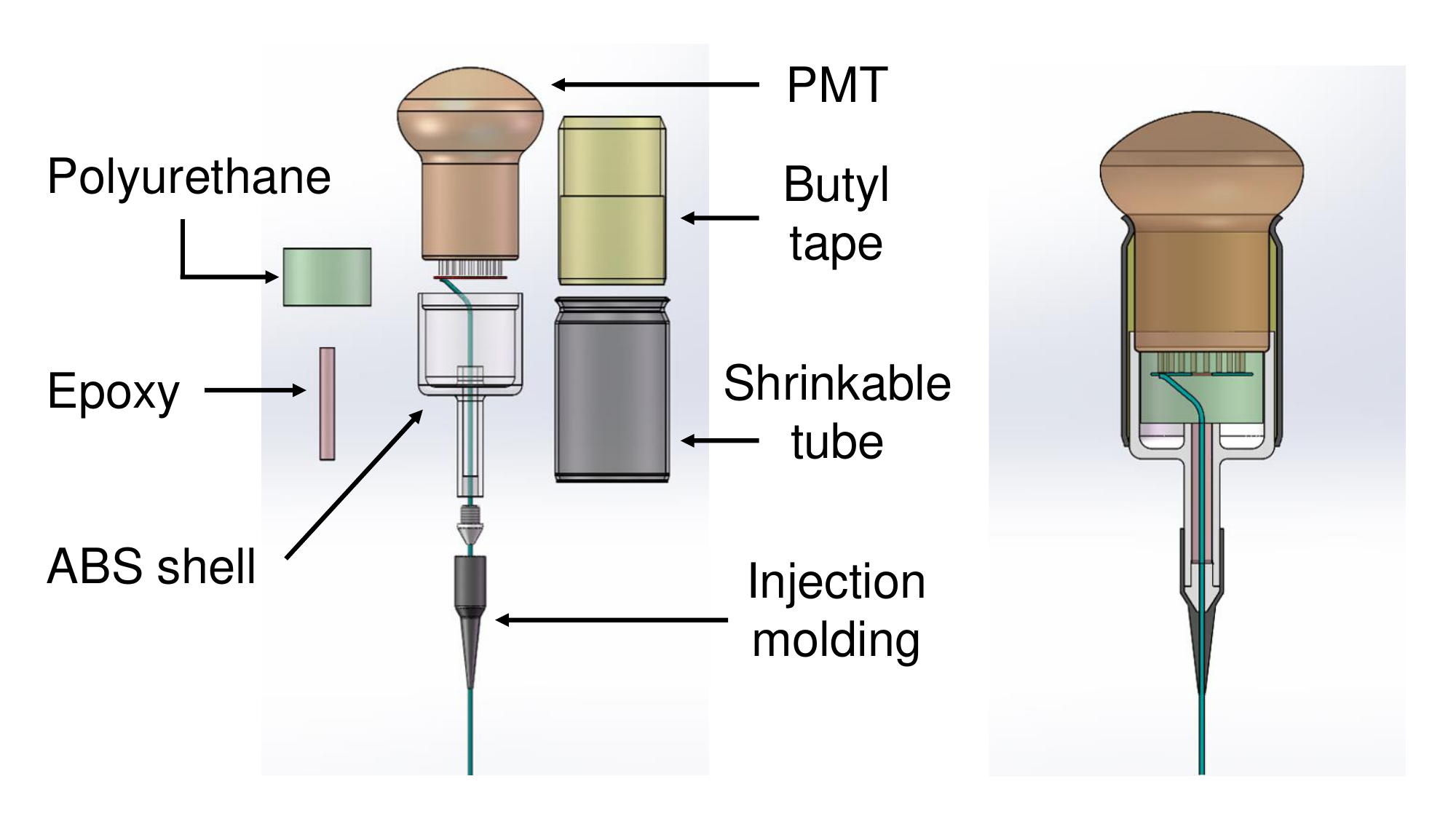}
  \caption{Schematic design of waterproof PMT potting.}
  \label{fig:potting}
\end{figure}

The schematic design of the waterproof PMT potting is illustrated in Fig.~\ref{fig:potting}. The PMT pins, the HV divider, and the cable are soldered and enclosed in an Acrylonitrile Butadiene Styrene (ABS) plastic shell. This shell is filled with polyurethane to protect the electrical components from water and to block any light that might be emitted spontaneously by electrical discharges. The seal between the shell and the PMT glass is achieved using butyl tape, which is encompassed by a shrinkable tube and finished with low-pressure injection molding at the end.

As a final validation of the waterproof potting design, 50 samples were fabricated using PMT glass bulbs or discarded PMTs, in combination with HV dividers and short cables. The cable terminations were sealed with epoxy, and the samples were immersed in a water tank in four separate batches, and subjected to a pressure of 0.6~MPa for a minimum of seven days. A voltage difference of 2,000~V was applied across the cable terminations by cutting the cable ends and removing the epoxy seal. The currents were recorded both before and after the high-pressure water test. For the glass bulb samples, the current measurements remained consistent. For the still functional discarded PMT samples, variations in current levels were noted, ranging from -40\% to 70\%, which could be attributed to differences in PMT light exposure. If water had infiltrated the PMT pins or the divider, a substantial increase in current, at least by an order of magnitude, would have been detected. Consequently, no instances of water leakage were detected in any of the tested samples.

\begin{figure}[!hbt]
\centering
\begin{subfigure}{.71\textwidth}
  \centering
  \includegraphics[width=.95\textwidth]{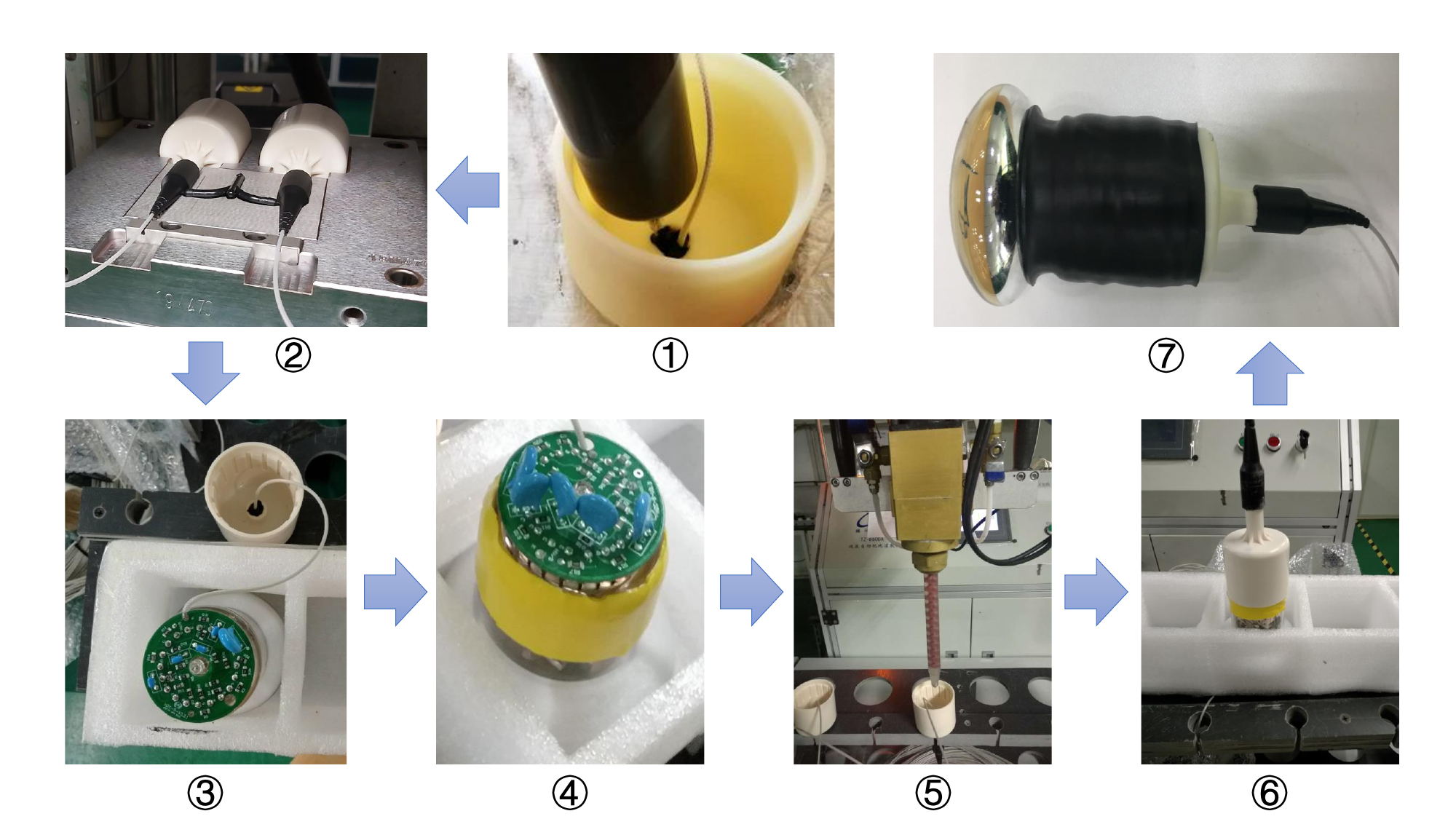}
  \caption{}
\end{subfigure}
\begin{subfigure}{.2\textwidth}
  \centering
  \includegraphics[width=.95\textwidth]{pmt_group2.png}
  \caption{}
\end{subfigure}
\caption{(a) Major steps of waterproof PMT potting: 1, cable sealing with epoxy; 2, cable sealing with injection molding; 3, soldering HV divider on the bare PMT; 4, assembly of butyl tape around the PMT; 5, filling polyurethane; 6, inverting the PMT; 7, covering with  shrinkable tube. The PMT signals were verified after step 3 and step 7 with HV. (b) A group of 16 PMTs with cables and the connector after waterproof potting.}
\label{fig:procedure}
\end{figure}

The primary steps involved in the potting are illustrated in Fig.~\ref{fig:procedure}~(a). An injection-molded ABS plastic potting shell was designed to provide mechanical support. Initially, the cable was threaded through a bottom entrance with a diameter of 2.2~mm, ensuring 28~cm of the cable remained inside the shell. Following this, epoxy Loctite ES4412~\cite{loctite} was injected from the top to offer a bonding length greater than 8~cm, serving as the inner seal. The epoxy used is specialized for HDPE, featuring a bonding strength of 1.6~MPa. The sealing was further reinforced outside the shell's tail through low-pressure injection molding with a polyamide and polyolefin mixture, also customized for HDPE.

Subsequently, the PMT was mounted and soldered onto the HV divider along with the cable. A set of 16 PMTs with plugs were then transferred to a dark box, where each channel was connected to a splitter box to separate the HV and the signal. A nominal HV was applied to each channel, and the current was monitored using a micro-ammeter. Utilizing low-intensity LED lights inside the dark box, PMT signals, predominantly SPEs, were captured by an oscilloscope. The amplitude was required to be within $\pm$50\% of the average. If a significant deviation was detected, the PMT was replaced, and the process was repeated for a new PMT.

A fixed volume of a two-part polyurethane adhesive was mixed and filled into the potting shell. The adhesive did not completely fill the potting shell, leaving a few milliliters of air to allow for sealant expansion due to temperature changes. The PMT glass was encircled by a 5~cm wide layer of butyl tape near the bottom of the PMT and positioned in the shell, where a pedestal provided mechanical support. This butyl tape filled the gap between the PMT glass and the shell. The PMT with the potting shell was inverted to ensure complete coverage of all electrical components by the polyurethane, and it was left for 24~hours to allow the polyurethane to cure. A second layer of butyl tape was then added outside the shell as an outer seal, with a 2~cm overlap on the first layer. Finally, a plastic shrinkable tube was added to protect the layers of butyl tape. The fully potted PMTs were tested again to verify signal quality. A picture of an integrated group of 16 PMTs is shown in Fig.~\ref{fig:procedure}~(b).

\subsection{Leakage Test}

\begin{figure}[ht]
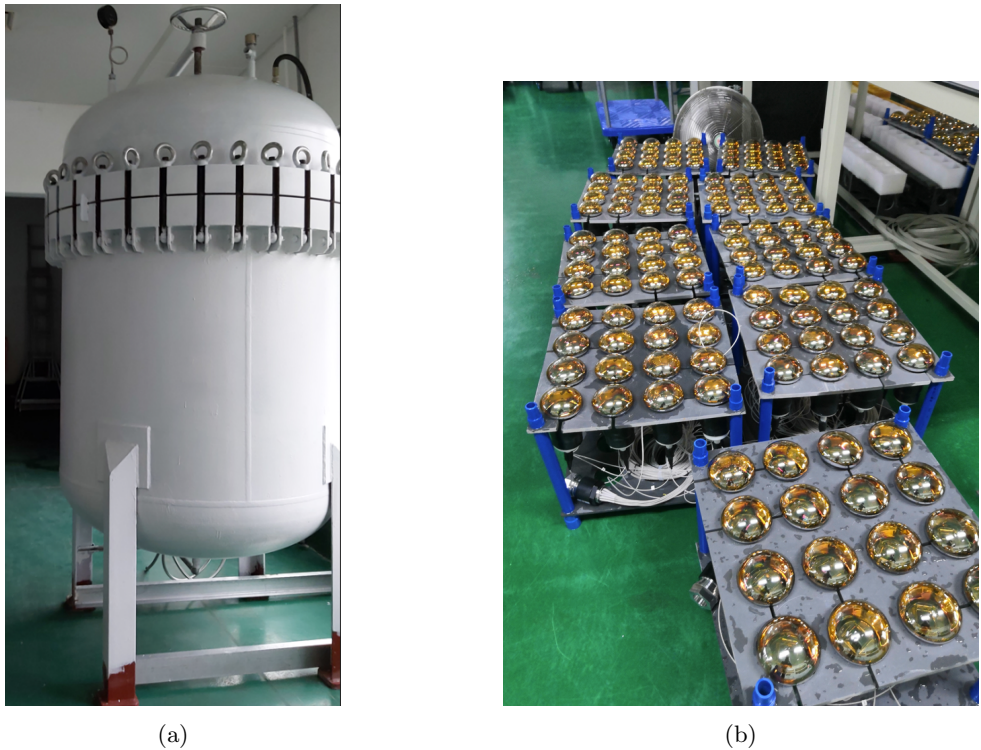

\centering
\begin{subfigure}{.45\textwidth}
  \centering
  \includegraphics[width=.6\textwidth]{HZCtank.png}
  \caption{}
\end{subfigure}
\begin{subfigure}{.45\textwidth}
  \centering
  \includegraphics[width=.85\textwidth]{HZCLeakageTest.png}
  \caption{}
\end{subfigure}
\caption{(a) Water tank used for the PMT potting leak testing. (b) Nine groups of PMTs taken out of the tank after 48 hours. Each group of 16 PMTs was fixed on a plastic rack.}
\label{fig:hzcWaterPressure}
\end{figure}

A large steel water tank was utilized to conduct a sampling test on potted PMTs to verify water leakage, as depicted in Fig.~\ref{fig:hzcWaterPressure}. A group of 16 PMTs was secured on a rack, with the plug sealed using two O-rings within a customized receptacle. Nine groups of PMTs of identical weight classification were simultaneously placed in the tank. The relative water pressure was elevated according to the weight specifications listed in Table~\ref{tab:wgroup} and maintained for 48~hours. Subsequently, the PMTs were removed from the tank, rinsed with pure water, and dried in an oven at 55$^{o}$C. The PMTs were then subjected to HV testing, and signals were measured for the third time in the dark box.
 
A total of 217 groups of PMTs, constituting approximately 13\% of the entire collection, were randomly selected from all three weight classes and subjected to underwater testing.
 
No potting leaks were detected in any of the groups. Nevertheless, four PMT glasses fractured during the water pressure test. Of these, two belonged to weight class W1, and the other two to weight class W2. The failure rate was approximately 0.1\%. These PMTs were subsequently replaced.

\section{Acceptance Test of Potted PMTs}
\label{sec.testing}

After potting, all PMTs underwent another round of testing to ensure their reliability and performance prior to being delivered to the JUNO site. This section provides details of the acceptance tests conducted.

\subsection{Test Station}
\label{subsection-Test station}

\begin{figure}[!hbt]
  \centering  
  \includegraphics[height=0.45\textwidth]{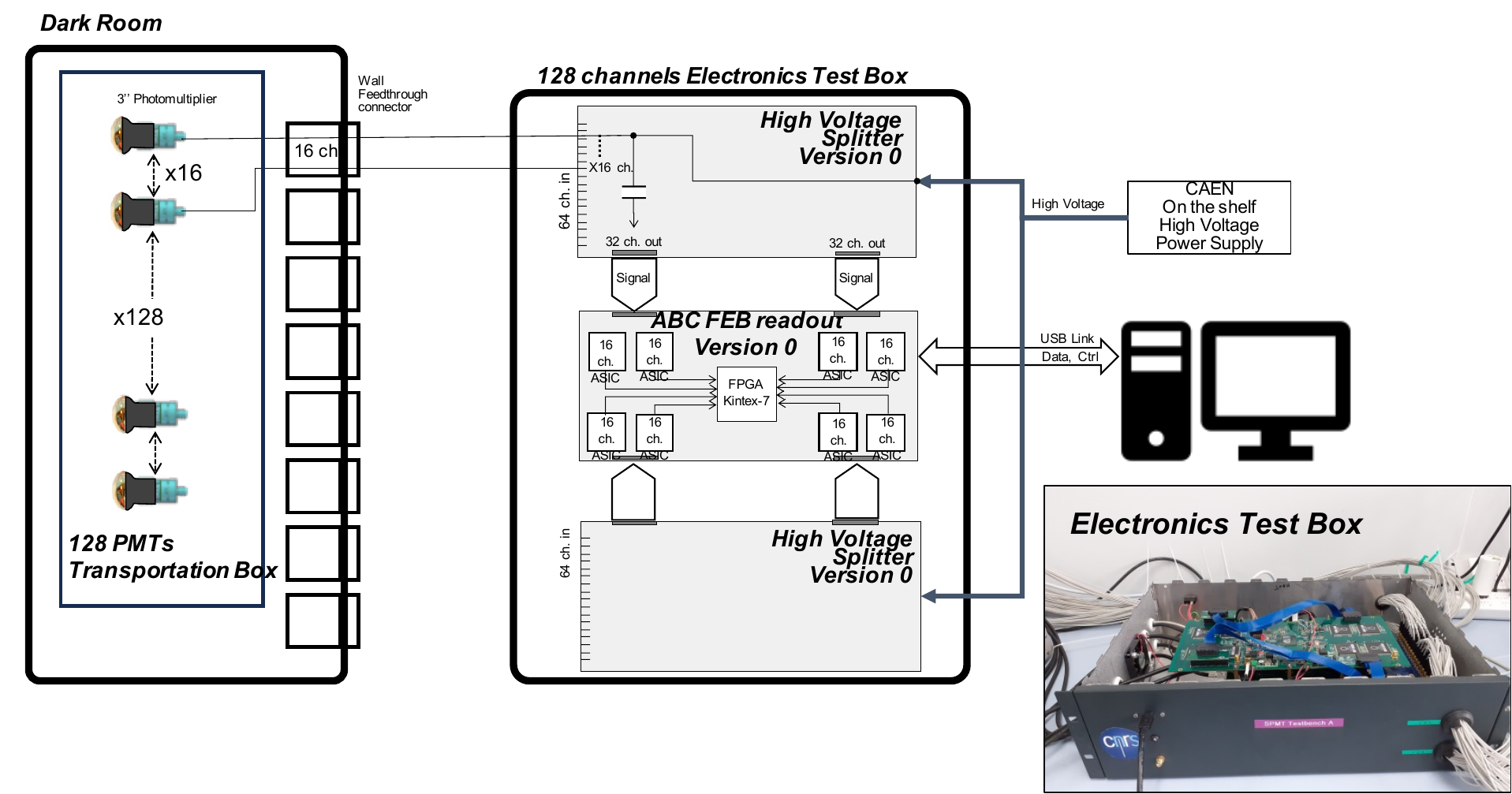}
  \caption{Schematics of a 128-channel PMT acceptance test station and an image of a 128-channel electronics test box.}
  \label{fig:acceptanceteststation}
\end{figure}

To ensure efficiency and speed during testing, a test station consisting of two test benches was constructed at Guangxi University (GXU). Each test bench can simultaneously handle eight groups of PMTs. The acceptance test setup included a dark room where the PMT transportation and storage boxes can be placed. Without removing the PMTs from their boxes, they can be connected via feedthrough wall connectors to a 128-channel electronic test box. Figure~\ref{fig:acceptanceteststation} provides a schematic view of the complete acceptance test system and a photo of the electronics box. This test box is an early simplified prototype of the electronics system. It comprised a 128-channel front-end readout board called the ABC (ASIC Battery Card), connected to the photomultipliers through two 64-channel splitter boards~\cite{Walker:2025xfa}. The prototype ABC board was equipped with 8 CATIROC ASIC chips~\cite{JUNO:2020orn}. Each prototype splitter board was designed to receive four HV inputs from a commercial HV power supply while still handling 64 channels. The firmware and software were customized for the acceptance test scenarios.

\subsection{Acceptance Criteria}

As stated in Sec.~\ref{sec.intro}, the bare PMTs were characterized in HZC.
The primary objective of the acceptance tests was to verify the functionality of each PMT following waterproof potting. Consequently, three main performance parameters were established as acceptance criteria, based on experimental requirements detailed in Table~\ref{tab-criteria}. The consistency of the working HV is a crucial parameter because groups of potted PMTs share the same HV. When the average gain for each group of PMTs is $3\times10^6$, the gain of each PMT should should range between  $2\times10^6$ and  $5\times10^6$. The SPE resolution of each PMT must be less than 45\%, and the dark count rate (DCR) must be lower than 3~kHz. 

	\begin{table}[!htb]
	\centering
	\caption{Acceptance test criteria.}
	\label{tab-criteria}
	\begin{tabular}{cc}
			\hline
			\hline
			Parameters & Requirements     \\ \hline
			Gain & $[2, 5] \times 10^6$      \\ 
			SPE resolution & $< 45\%$    \\ 
			Dark rate & $< 3$ kHz \\ 
			\hline
			\hline
	\end{tabular}
	\end{table}

\subsection{HV Calibration Testing}

\begin{figure}[!htbp]
\centering
\begin{subfigure}{.45\textwidth}
  \centering
  \includegraphics[width=.95\textwidth]{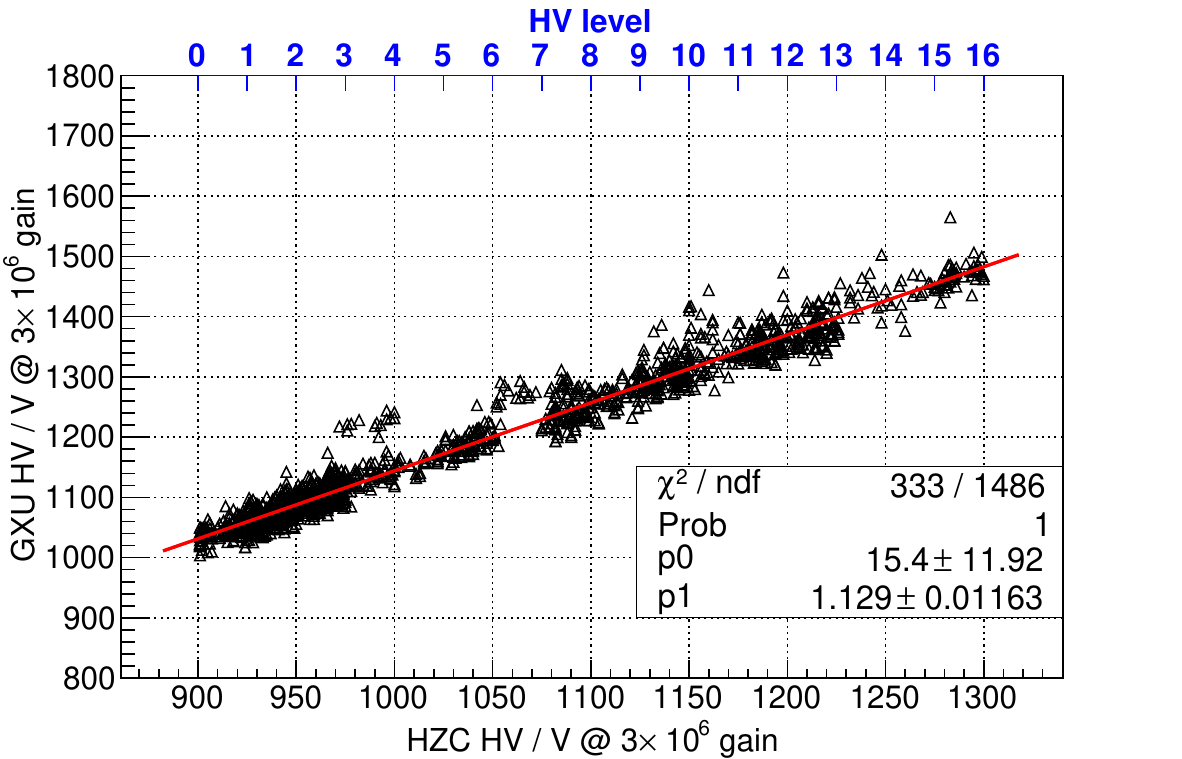}
  \caption{}
\end{subfigure}
\begin{subfigure}{.45\textwidth}
  \centering
  \includegraphics[width=.95\textwidth]{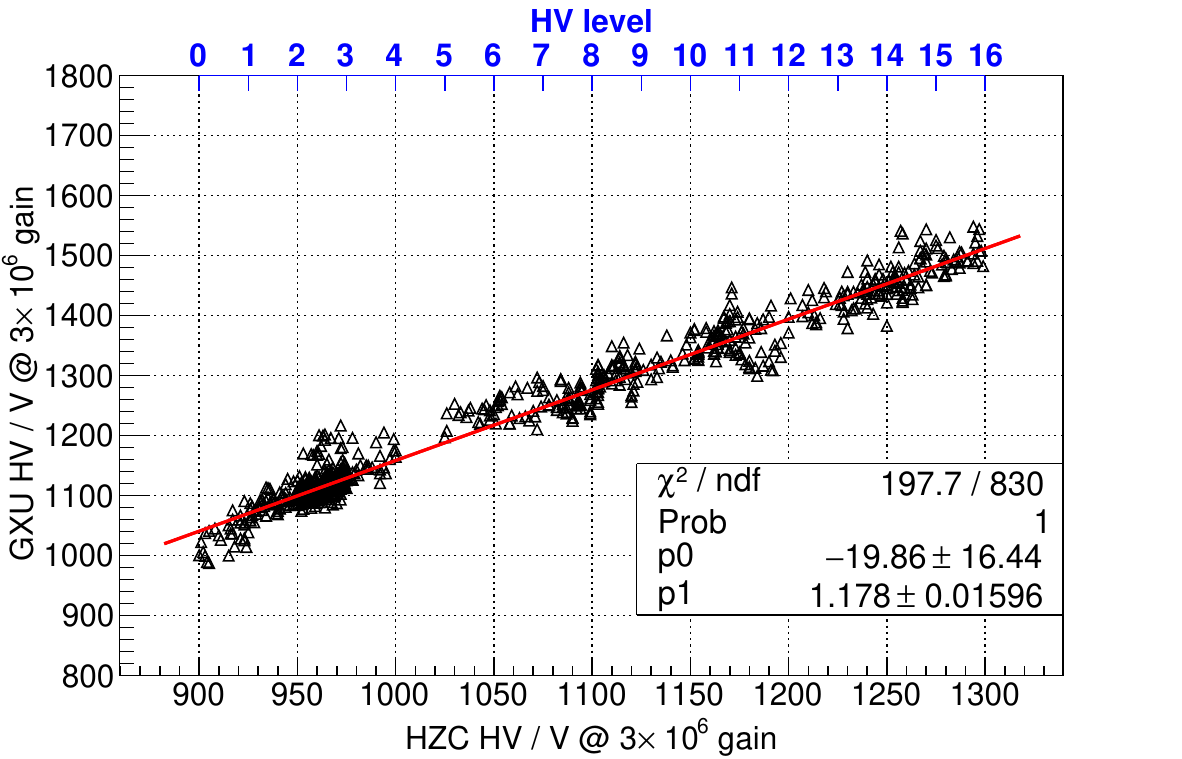}
  \caption{}
\end{subfigure}
\caption{Correlations between GXU and HZC HV measurements. The red line is the result of a simple linear fit to the data. (a) The cable length is 5 meters. (b) The cable length is 10 meters.}
\label{fig:HV-calib}
\end{figure}

The working HV for the potted PMTs is higher than that of the bare tubes. This is because, firstly, the HV for potted PMT is supplied through an HV splitter.  Each channel's 20~M$\Omega$ current-limiting resistor reduces the HV available to the tube by approximately 10\%. Secondly, as discussed in Sec.~\ref{sec.signal}, any attenuation in the long cable is compensated by applying a slightly higher HV. At the start of the acceptance test, the working HV for achieving a gain of $3\times10^6$ (designated as GXU HV) was determined  by systematically scanning various HV levels for small batches of PMTs. 
This value was then compared with the working HV determined from bare tube testing by the PMT manufacturer HZC (referred to HZC HV, with each increment of 25~V representing a distinct HV level). This comparison resulted in a linear relationship, as depicted in Fig.~\ref{fig:HV-calib}. 

Utilizing this correlation, the working HV for each group of PMTs can be easily derived from their HZC HV values during acceptance testing. Subsequent tests confirmed the effectiveness of this procedure, as only a small number of PMT groups required further adjustments to their working HV during testing.  These instances were often due to the challenge of assembling 16 PMTs with identical HV levels into a complete set during waterproof potting, resulting in PMTs with significantly different HV levels being included in the same group. Once all PMTs are installed in the JUNO detector, the scanning will be conducted again for every group to determine the precise working HV.

\subsection{Test Results}

As stated in Sec.~\ref{subsection-Test station}, for the efficient batch acceptance testing of 26,000 PMTs, we employed two test benches capable of simultaneously measuring 128 PMTs per test setup. The PMTs were placed inside a storage box and shielded in a dark room with a black cloth. It was found that using PMT dark noise hits for gain and SPE resolution measurements was the optimal choice. Before collecting test data, the PMTs were placed in the dark room and powered at their working HV for more than eight hours to reduce and stabilize the DCRs. Forced trigger mode (pedestal mode) was used in the tests to establish the baseline level of each channel. Subsequently, self-triggered mode was utilized to capture the SPE signals generated by PMT dark count events, with a trigger threshold set at 870 digital to analog converter units (DACu), corresponding to about 0.3~PE.

\begin{figure}[!hbt]
  \centering
  \includegraphics[height=0.65\textwidth]{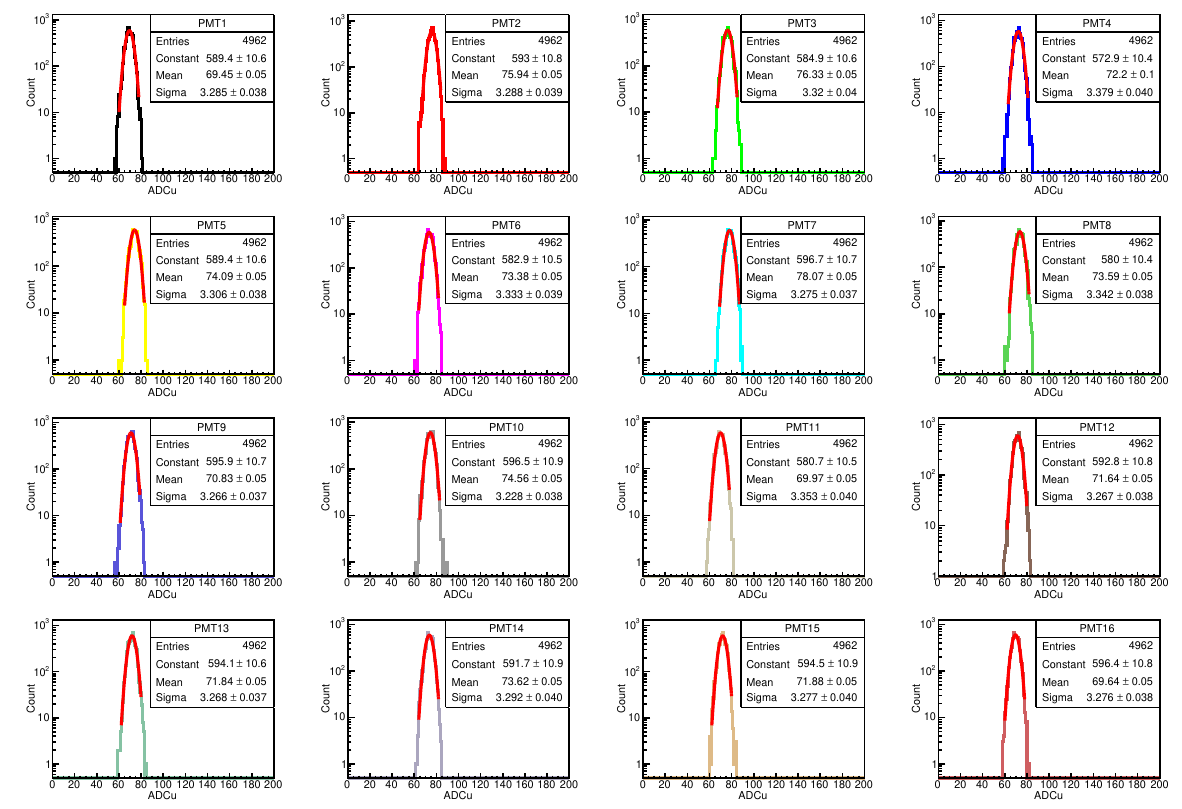}
  \caption{Pedestal charge spectra of a group of PMT.}
  \label{fig:Pedestal-spectrum}
\end{figure}

\begin{figure}[!hbt]
  \centering  
  \includegraphics[height=0.65\textwidth]{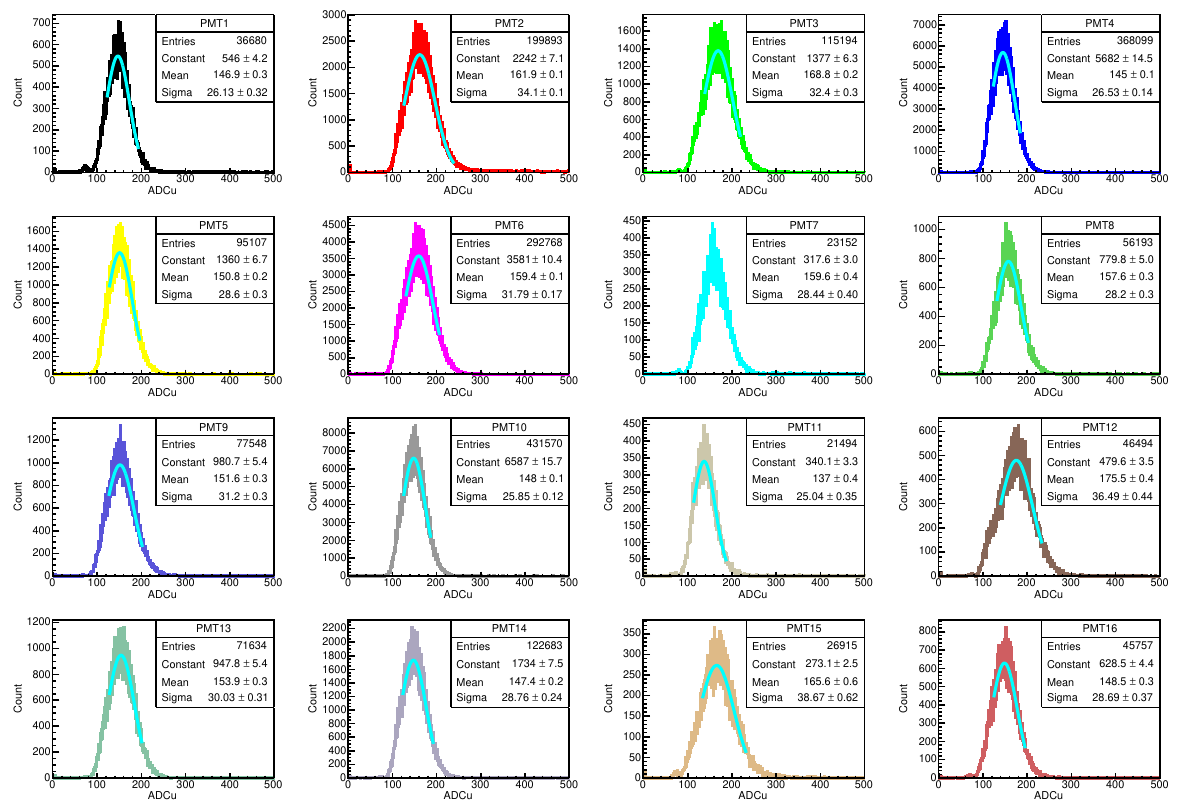}
  \caption{The SPE charge spectra of a group of PMTs.}
  \label{fig:spe-spectrum}
\end{figure}

Figure~\ref{fig:Pedestal-spectrum} and Fig.~\ref{fig:spe-spectrum} show the pedestal and SPE distributions for a group of 16 PMTs, respectively. By fitting these distributions with Gaussian functions, the analog to digital converter units (ADCu) corresponding to their peak positions can be determined, referred to as $Q_0$ and $Q_1$, respectively. Additionally, the sigma value of the Gaussian function is denoted as $\sigma_1$ for the SPE distribution. Consequently, the gain of each PMT can be calculated using the following formula:

\begin{equation}
\label{eq:gain}
{\rm Gain} = \frac{Q_1-Q_0}{1.6 \times 10^{-7} \times P_1},
\end{equation}
where $P_1$ is the calibration constant of each 
CATIROC channel, which is $160 \sim 190$ ADCu/pC for a CATIROC preamplifier gain of 40. As a result, the estimated range for the $\Delta$ADC value ($Q_1 - Q_0$) corresponding to a $3 \times 10^6$ gain is approximately from 77 ADCu to 91 ADCu.

In the analysis of the test data, to facilitate the comparison and evaluation of gain consistency within the same group of PMTs under a specified working HV, the average gain of the group was normalized to $3 \times 10^6$. The gains of PMTs in the same group were scaled by the same normalization factor. Subsequently, PMTs that did not meet the acceptance criteria (gain below $2\times10^6$ or above $5\times10^6$) were excluded and replaced with spare PMTs that aligned with the working HV.

\begin{figure}[!htb]
\centering
\begin{subfigure}{.45\textwidth}
  \centering
  \includegraphics[width=.95\textwidth]{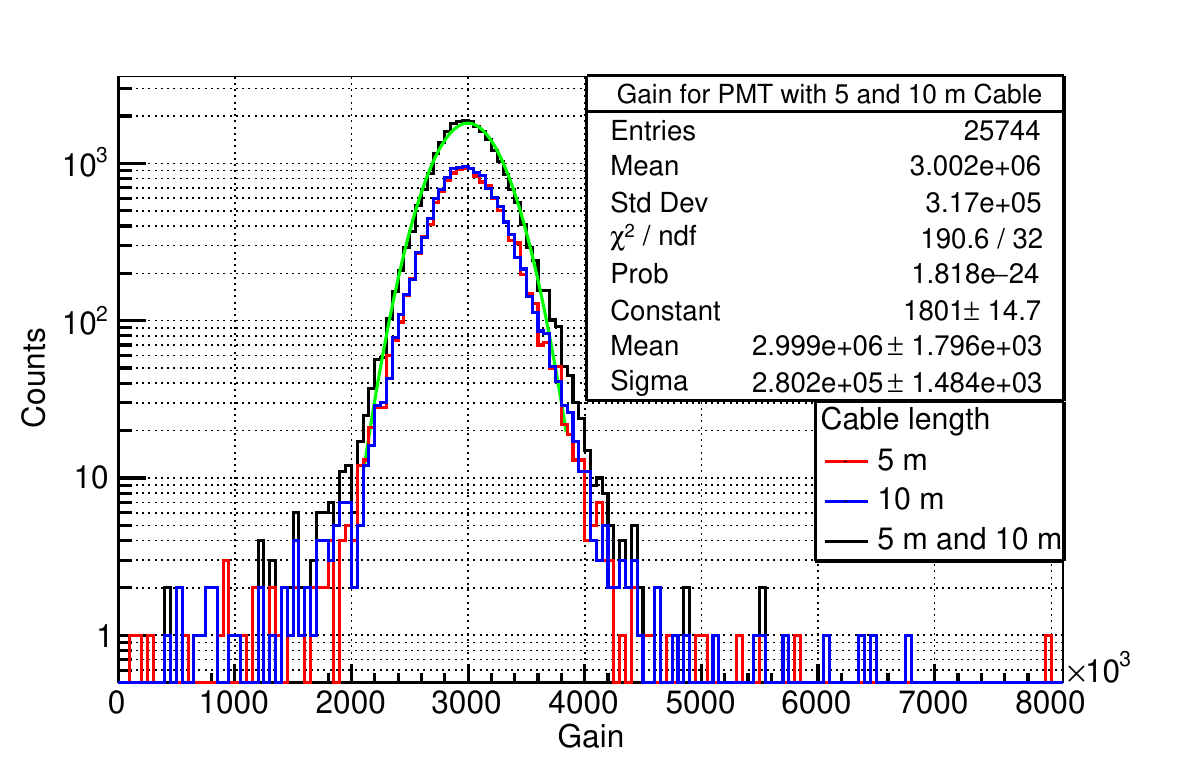}
  \caption{}
  \label{fig:gain-spread-1}
\end{subfigure}
\begin{subfigure}{.45\textwidth}
  \centering
  \includegraphics[width=.95\textwidth]{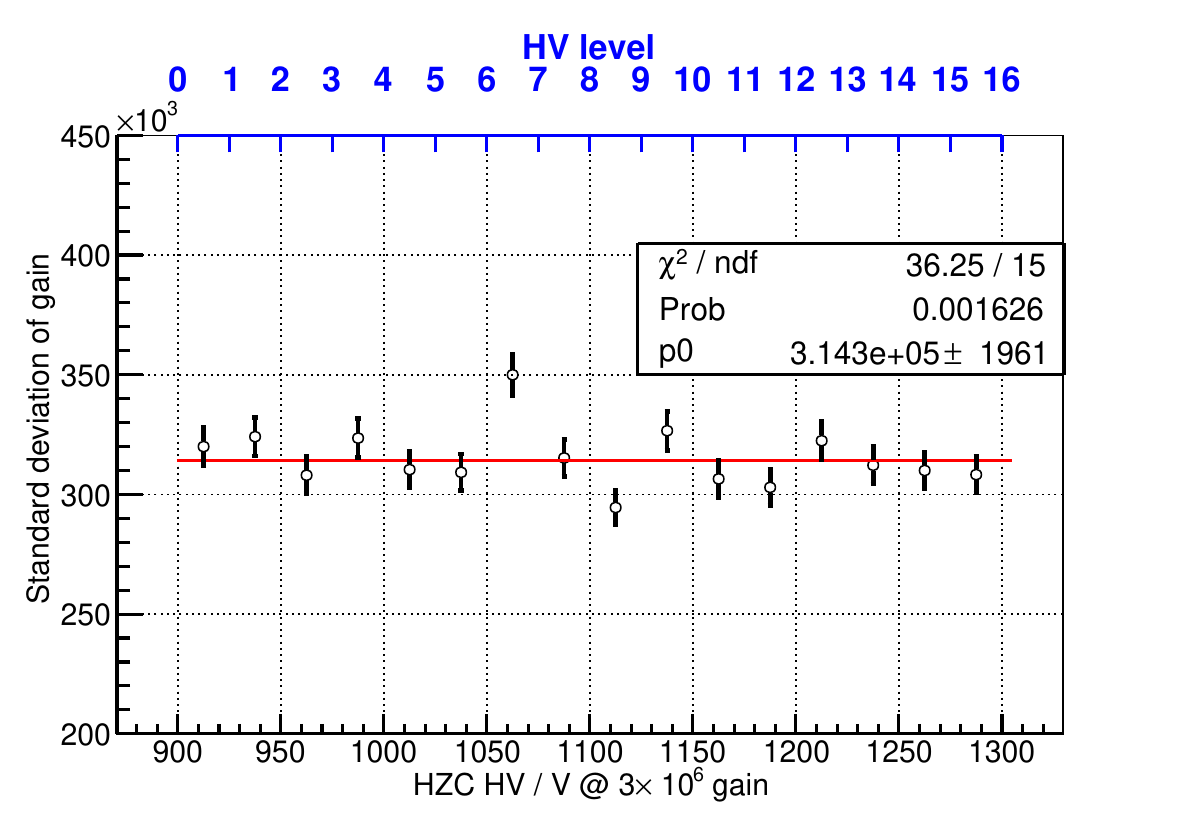}
  \caption{}
  \label{fig:gain-spread-2}
\end{subfigure}
\caption{(a) Gain distribution of the measured PMTs. (b) Gain spread of PMTs across the 16 HV levels. The red line represents a constant fit to all points, illustrating that the spread remains consistently around $\sim$10\% across all HV levels.}
\label{fig:gain-spread}
\end{figure}

The gain distribution of the measured PMTs as obtained from this analysis is shown in Fig.~\ref{fig:gain-spread-1}. The results presented in Fig.~\ref{fig:gain-spread-2} demonstrate that the gain spread remains consistently around $\sim10\%$ across different HV levels. 

In addition, the SPE resolution was obtained as follows:

\begin{equation}
\label{eq:reso}
{\rm SPE~resolution (\%)}= \frac{\sigma_1}{Q_1-Q_0} \times 100\%
\end{equation}

\begin{figure}[!htb]
\centering
\begin{subfigure}{.45\textwidth}
  \centering
  \includegraphics[width=.95\textwidth]{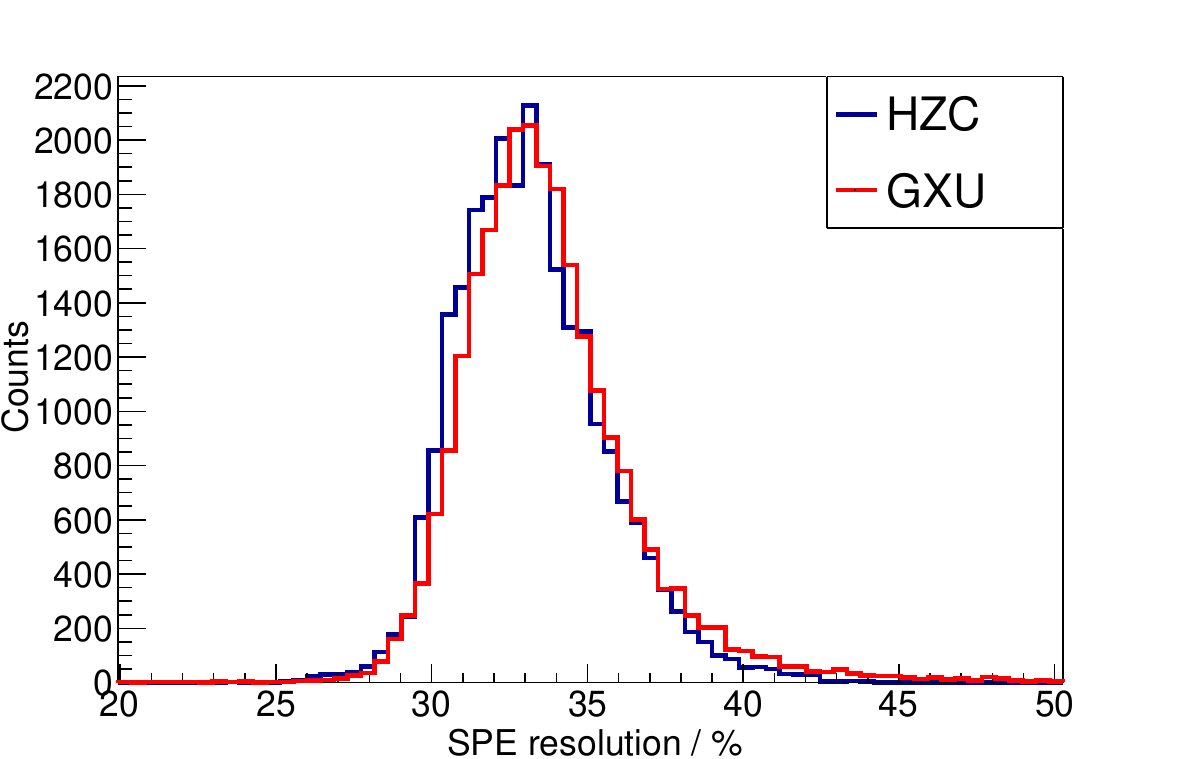}
  \caption{}
  \label{fig:resolution-1}
\end{subfigure}
\begin{subfigure}{.45\textwidth}
  \centering
  \includegraphics[width=.95\textwidth]{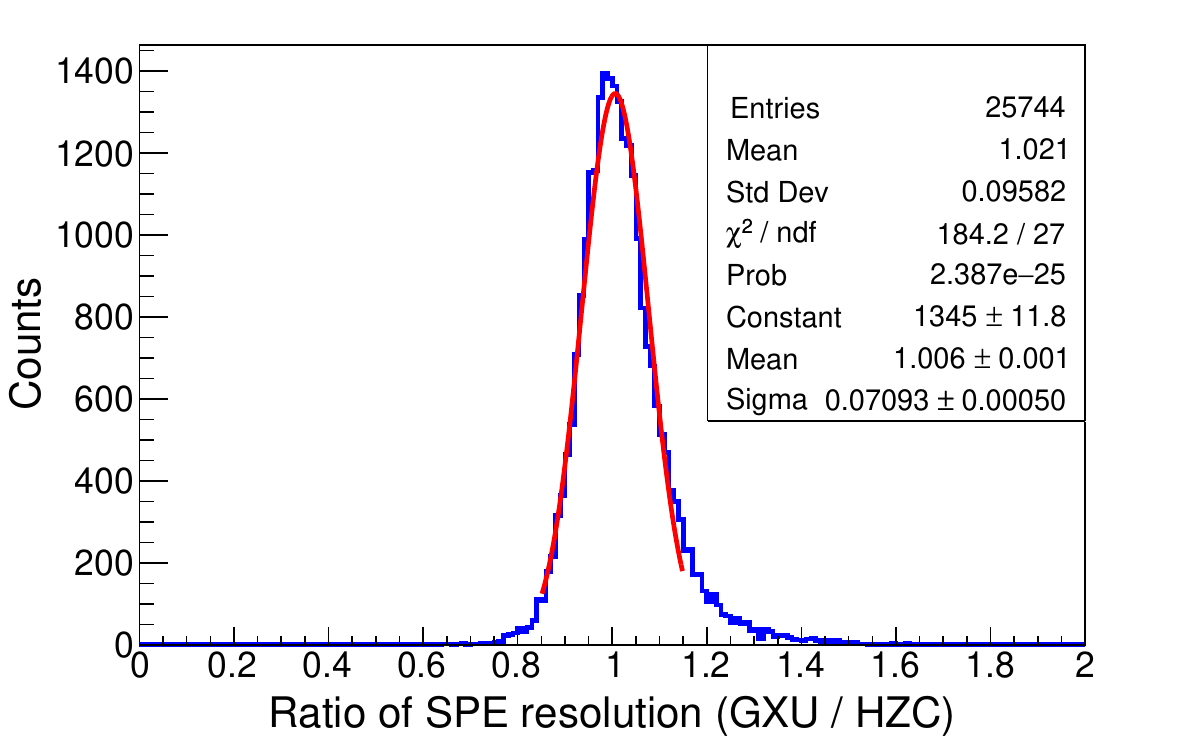}
  \caption{}
  \label{fig:resolution-2}
\end{subfigure}
\caption{(a) The SPE resolution of the measured PMTs. (b) The comparison of SPE resolution between the potted PMTs and the bare PMTs.}
\label{fig:resolution}
\end{figure}

The SPE resolution results are presented in Fig.~\ref{fig:resolution}, indicating that the majority of PMTs exhibit a value below 45\%, thereby satisfying the acceptance criteria. The SPE resolution was further compared with the results of the bare tube tests, and good consistency was observed.

\begin{figure}[!hbt]
  \centering  
  \includegraphics[height=0.35\textwidth]{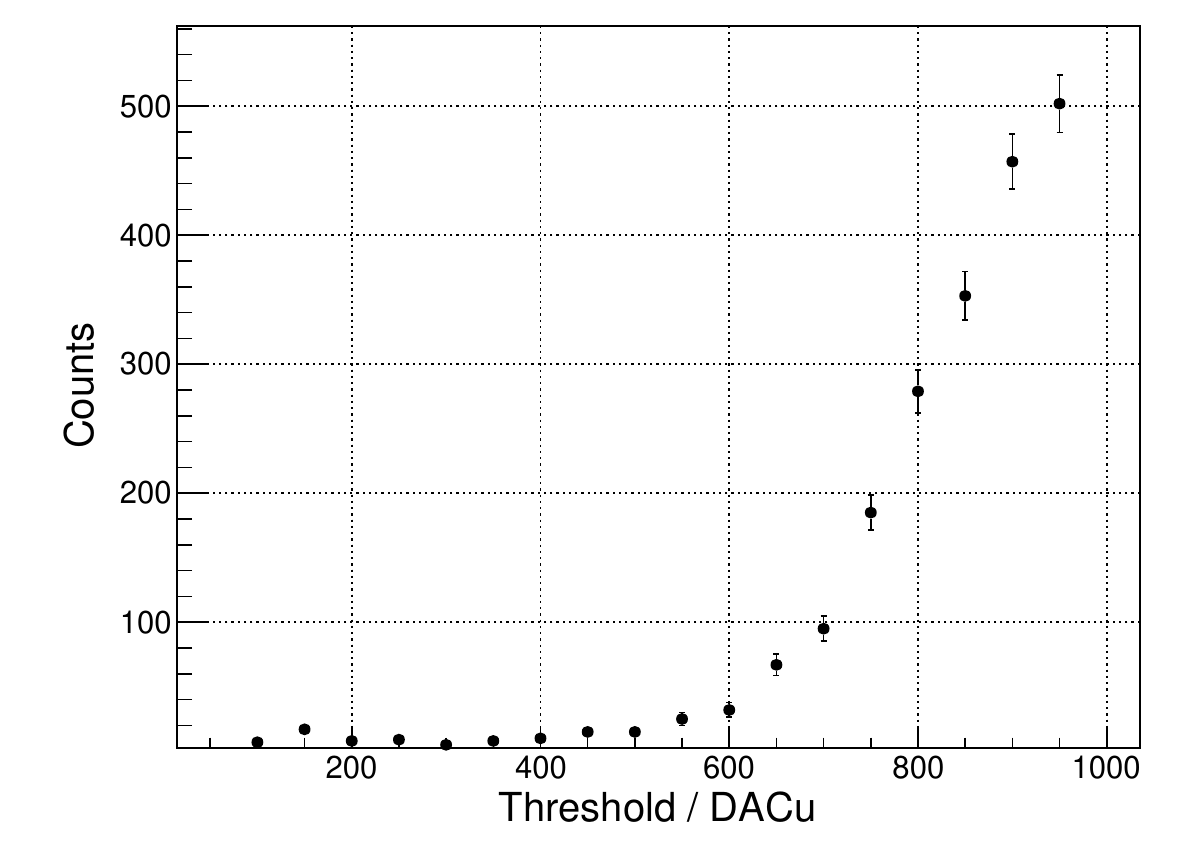}
  \caption{The S-curve for one PMT. The number of triggers counted in 0.419~s as a funnction of the trigger threshold in DAC unit (a high DACu value corresponds to a low trigger threshold value in amplitude).
  }
  \label{fig:HV.darkrateScan}
\end{figure}

\begin{figure}[!hbt]
  \centering  
  \includegraphics[height=0.3\textwidth]{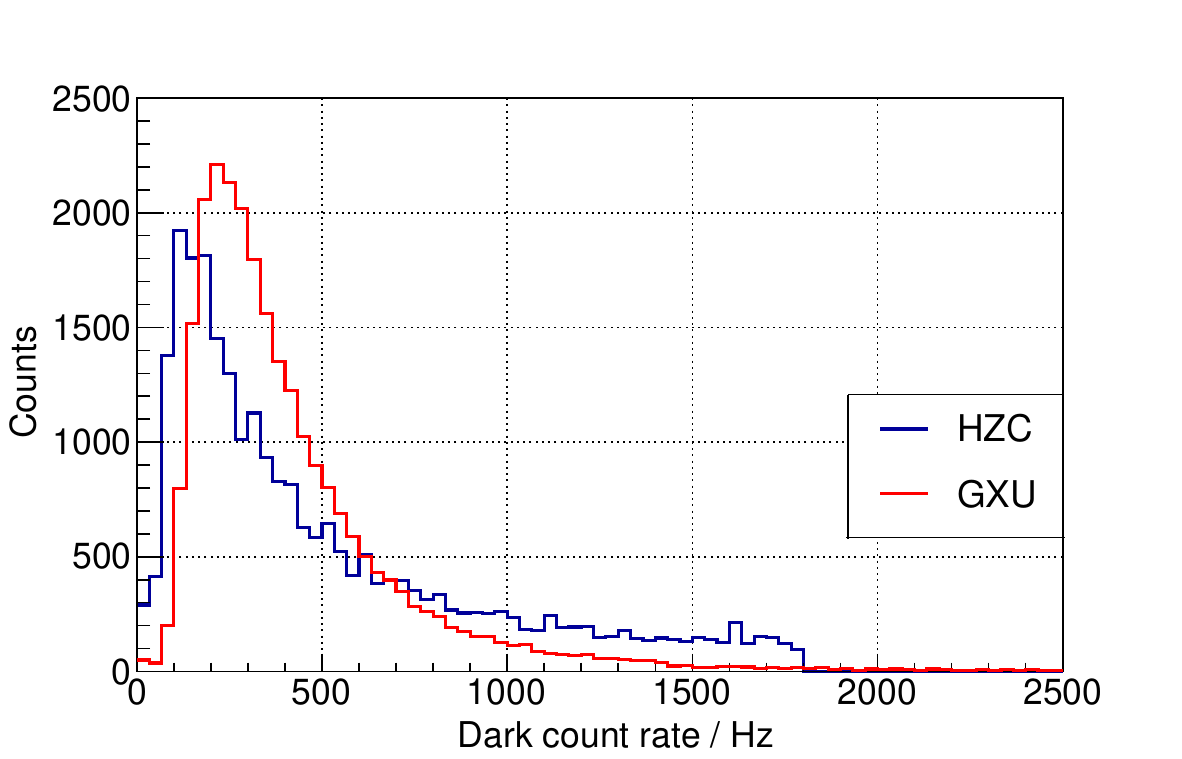}
  \caption{PMT dark rate distributions measured at the company (HZC) and during the acceptance tests at Guangxi University (GXU). }
  \label{fig:HV.darkrate}
\end{figure}

Another important acceptance parameter is the DCR, which is measured by the threshold scanning method. In this test, the number of trigger signals from each PMT channel in an electronics time period of 0.419~s is measured as a function of the trigger threshold in DAC units ranging from 0 to 870. The trigger values are in inverse order: a high DACu value corresponds to a low trigger threshold value in amplitude. The counts versus the discriminator threshold (also called the S-curve) is shown in Fig.~\ref{fig:HV.darkrateScan}. The PMT DCR was obtained by calculating the trigger rate at threshold 870 DACu, which corresponds to $\sim$ 0.3 PE. Figure~\ref{fig:HV.darkrate} shows the DCR distribution of the potted PMTs and the bare PMTs. The average DCR for the bare PMTs is approximately 510~Hz, with a significant tail in the distribution primarily attributed to insufficient time spent in the dark room under high voltage (HV) for the DCR to stabilize before testing. For instance, some PMTs were only subjected to a 2-hour HV load, and a DCR value below 3~kHz was considered acceptable. The potted PMTs benefited from sufficient cooling time with working HV of 8 hours, resulting in a significant reduction in the tail component and an average dark count rate of approximately 420~Hz, lower than that observed for the bare PMT test in HZC. However, upon comparing the main peak position of the dark count rate distribution, a significant shift is observed in the acceptance test, with the peak position approximately 100~Hz higher than that in the bare PMT test. Our investigation reveals that this additional contribution primarily comes from fluorescence emitted by the packaging foam employed to protect the PMTs. Subsequent tests conducted on small batches of PMTs demonstrated a difference of around 100~Hz in dark count rates between foam-protected and non-foam conditions. Given that the primary objective of the acceptance test is to validate the functionality of the potted PMTs, it was deemed unnecessary to conduct extensive testing without foam protection in order to ensure efficiency and prevent any damage during this process.

In total, 25,744 instrumented PMTs, corresponding to 1,609 groups, were tested. The rejected ratio of the acceptance test is about 0.7\%. Table~\ref{tab-result} summarizes the number of unqualified PMTs and the issues identified during testing. PMTs that did not pass the tests were removed and replaced with spare PMTs. The replaced PMTs were subsequently tested at the JUNO site, where they all met the acceptance criteria.

	\begin{table}[htbp]
	\centering
	\caption{Results of the acceptance tests for specific criteria, based on a total of 25,744 tested PMTs.}
	\label{tab-result}
	\begin{tabular}{cc}
			\hline
			\hline
			Issue & Unqualified num.    \\ \hline
			Gain + Resolution + Dark count rate & 1  \\
			Gain + Resolution & 14  \\ 
			Gain + Dark count rate & 0  \\ 
			Resolution + Dark count rate & 0 \\ 
			Gain & 58     \\ 
			SPE resolution & 14    \\ 
			Dark rate & 81  \\ 
			Short-circuit & 9 \\ \hline
			Total & 177  \\      
            \hline
                \hline
	\end{tabular} 
	\end{table}

\section{Summary}
\label{sec.summary}
The instrumentation of the 3-inch PMTs of JUNO underwent rigorous testing at every stage of development and mass production. Each PMT was equipped with a divider capable of supplying up to 1,300~V across the various dynodes. Capacitors with wires and surface-mounted resistors were meticulously chosen to meet space constraints while ensuring reliability. A standard RG178 coaxial cable, available in 5~m or 10~m lengths, was used to transmit both HV and the signal. The cable jacket was customized to enhance waterproofing. A 16-channel waterproof connector, developed in collaboration with the Axon' cable company, served as the interface between the PMTs and the electronics. Signal attenuation in the cable was adequately compensated by the voltage, and the observed analog crosstalk was below 0.3\%. Two long-term underwater validations were conducted, demonstrating promising reliability with FIT values of less than 500 and 300, respectively, in line with JUNO's requirements.

To integrate PMTs with HV dividers, cables, and connectors, the PMTs were grouped into more than 1,600 sets based on their weights and operating HV. Each group was connected to a single connector and arranged in the water pool from shallow to deep, according to weight, from lightest to heaviest. Waterproof potting of the PMTs was achieved through multiple redundant layers of sealants with mechanical support. A random selection of 217 groups was tested underwater, and no leakage was found in the PMT sealing. 

All potted PMTs were tested with an early version of the electronics system. The SPE spectrum of each group was verified under the nominal threshold. A 10\% standard deviation from the nominal 3$\times$10$^{6}$ gain was found. After potting, the SPE resolution remained unchanged, and there was no significant variation in the DCR. Only 0.7\% of potted PMTs were replaced due to high DCRs, significant deviations from the target gain, poor charge resolution, or electrical short circuits. Altogether, 25,600 instrumented PMTs, along with an additional 144 spares, were delivered to JUNO.

\section{Acknowledgement}

We are grateful for the ongoing cooperation from the China General Nuclear Power Group.
This work was supported by
the Chinese Academy of Sciences,
the National Key R\&D Program of China,
the Guangdong provincial government,
and the Tsung-Dao Lee Institute of Shanghai Jiao Tong University in China,
the Institut National de Physique Nucl\'eaire et de Physique de Particules (IN2P3) in France,
the Istituto Nazionale di Fisica Nucleare (INFN) in Italy,
the Fond de la Recherche Scientifique (F.R.S-FNRS) and the Institut Interuniversitaire des Sciences Nucl\'eaires (IISN) in Belgium,
the Conselho Nacional de Desenvolvimento Cient\'ifico e Tecnol\`ogico in Brazil,
the Agencia Nacional de Investigacion y Desarrollo and ANID - Millennium Science Initiative Program - ICN2019\_044 in Chile,
the European Structural and Investment Funds, the Czech Ministry of Education, Youth and Sports and the Charles University Research Centerin Czech Republic,
Deutsche Forschungsgemeinschaft (DFG), the Helmholtz Association, and the Cluster of Excellence PRISMA+ in Germany,
the Joint Institute of Nuclear Research (JINR) and Lomonosov Moscow State University in Russia,
MOST and MOE in Taipei,
the Program Management Unit for Human Resources \& Institutional Development, Research and Innovation, Chulalongkorn University, and Suranaree University of Technology in Thailand, 
the Science and Technology Facilities Council (STFC) in the UK, 
University of California at Irvine and the National Science Foundation in the US.

\bibliographystyle{h-physrev5}
\bibliography{references}

\end{document}